\documentclass[runningheads]{llncs}
\usepackage[utf8]{inputenc}
\usepackage{tikz}
\usepackage{tikz-qtree}
\usetikzlibrary{positioning}
\usetikzlibrary{trees}
\usetikzlibrary{decorations.pathreplacing,calc}
\usetikzlibrary{shapes.misc}
\usepackage{listings}
\usepackage{bm}
\usepackage{amsmath}
\usepackage{amssymb}
\usepackage{mathrsfs}
\usepackage{csquotes}

\let\subparagraph\paragraph

 \usepackage[compact]{titlesec}

\usepackage[subtle]{savetrees}

\usepackage[linewidth=0.5pt]{mdframed}
\usepackage[scaled=0.85]{beramono}
\usepackage{algorithm}
\usepackage{algpseudocode}
\usepackage{algorithmicx}
\usepackage[english]{babel}
\usepackage{biblatex}
\usepackage{graphicx}
\usepackage{setspace}
\usepackage{enumitem}
\usepackage{multicol}
\usepackage{verbatim}

\titlespacing{\subsubsection}{2pt}{1pt}{1pt}

\title{Pointer Data Structure Synthesis from Answer Set Programming Specifications}
\titlerunning{Pointer Data Structure Synthesis from ASP Specifications}
\author{}
\date{April 2020}

\newcommand{\mycomment}[1]{}
\newcommand{\pispec}{\Pi_{upd}}

\newcommand{\pisym}{\Pi^{sym}_{upd}}
\newcommand{\pisymbst}{\Pi^{sym}_{bst}}
\newcommand{\pisymprime}{\Pi^{sym'}_{upd}}
\newcommand{\pisymprimebst}{\Pi^{sym'}_{bst}}
\newcommand{\pisymprimedeux}{\Pi^{sym''}_{upd}}
\newcommand{\pifrag}{\Pi_{f}}
\newcommand{\pitrav}{\Pi_{trv}}
\newcommand{\pisymtrav}{\Pi^{sym}_{trv}}
\newcommand{\pisymtravprime}{\Pi^{sym'}_{trv}}
\newcommand{\pisymtravprimedeux}{\Pi^{sym''}_{trv}}
\newcommand{\piinput}{\Pi_{inp}}
\newcommand{\pieq}{\Pi_{eq}}

\urldef{\mailsa}\path|{sxv153030, neerajm, gupta}@utdallas.edu|

\begin{document}

\author{Sarat Chandra Varanasi
\and Neeraj Mittal \and Gopal Gupta}

\institute{Department of Computer Science,\\The University of Texas at Dallas,\\
Richardson, Texas, USA\\
\mailsa}

\setlength{\abovedisplayskip}{0pt}
\setlength{\belowdisplayskip}{0pt}
\setlength{\abovedisplayshortskip}{0pt}
\setlength{\belowdisplayshortskip}{0pt}

\newcommand{\tikzmark}[1]{\tikz[overlay,remember picture] \node (#1) {};}
\newcommand*{\AddNoteLeft}[4]{%
    \begin{tikzpicture}[overlay, remember picture]
        \draw [decoration={brace,amplitude=0.5em,mirror},decorate,thin,black]
            ($(#3)!([yshift=1.5ex]#1)!($(#3)-(0,1)$)$) --  
            ($(#3)!(#2)!($(#3)-(0,1)$)$)
                node [align=center, text width=2.5cm, pos=0.5,anchor=west] {#4};
    \end{tikzpicture}}

\newcommand*{\AddNoteRight}[4]{%
    \begin{tikzpicture}[overlay, remember picture]
        \draw [decoration={brace,amplitude=0.5em},decorate,thin,black]
            ($(#3)!([yshift=1.5ex]#1)!($(#3)-(0,1)$)$) --  
            ($(#3)!(#2)!($(#3)-(0,1)$)$)
                node [align=center, text width=2.5cm, pos=0.5,anchor=west] {#4};
    \end{tikzpicture}}

\maketitle

\begin{abstract}
 We develop an inductive proof-technique to generate imperative programs for pointer data structures from behavioural specifications expressed in the Answer Set Programming (ASP) formalism. ASP is a non-monotonic logic based formalism that employs\textit{negation-as-failure} which helps emulate the human thought process, allowing domain experts to model desired system behaviour succinctly. We argue in this paper that ASP's reliance on negation-as-failure makes it a better formalism than those based on first-order logic for writing formal specifications. 
 We assume the a domain expert provides the representation of inductively defined data structures along with a specification of its 
 operations. Our procedures combined with our novel proof-technique reason over the specifications and automatically generate an imperative program. Our proof-technique leverages the idea of partial deduction to simplify logical specifications. By algebraically simplifying logical specifications we arrive at a residual specification which can be interpreted as an appropriate imperative program. This work is in the realm of constructing programs that are correct according to a given specification.  
\end{abstract}

\lstdefinelanguage{Python}{
  %keywords={typeof, new, true, false, catch, function, return, null, catch, switch, var, if, in, while, do, else, case, break, def, for},
  keywordstyle=\color{black}\bfseries,
  ndkeywords={class, export, boolean, throw, implements, import, this},
  ndkeywordstyle=\color{darkgray}\bfseries,
  identifierstyle=\color{black},
  sensitive=false,
  comment=[l]{//},
  morecomment=[s]{/*}{*/},
  commentstyle=\color{purple}\ttfamily,
  stringstyle=\color{red}\ttfamily,
 % morestring=[b]',
 % morestring=[b]",
 % morestring=`[b]
}

\lstset{
   %language=JavaScript,
   backgroundcolor=\color{white},
   extendedchars=true,
   basicstyle=\footnotesize\ttfamily,
   showstringspaces=false,
   showspaces=false,
   numberstyle=\footnotesize,
   numbersep=9pt,
   tabsize=2,
   breaklines=true,
   showtabs=false,
   captionpos=b
}

   \lstset{%
       escapeinside={(*}{*)},%
      }
\section{Introduction}
     Declarative specifications allow a domain expert to model properties and behaviours of a system precisely. When specifications are executable, the expert can often test the specification quickly and do refinements when necessary. Executable specifications are often slow computationally and are non-deterministic in their search space exploration. Our work extracts deterministic imperative programs from executable specifications written in Answer Set Programming (ASP) by means of a novel proof-technique. Answer Set Programming \cite{gelfond2014knowledge} is a logical formalism based on non-monotonic logic and allows for succinct specifications of complex problems. The ASP formalism can specify a wide range of computational problems such as Graph Coloring, Planning and so on. Its primary use is in Artificial Intelligence to perform commonsense reasoning but can also express complex ideas involving dynamic properties of a system \cite{erdem2016applications, lifschitz2019answer, chen2016physician}. 
     
     %GG: paragraph added
     The main intuition behind our work is as follows: humans rely on non-monotonic reasoning in their day to day lives. That is, based on their current state of knowledge, humans jump to conclusions. Later, if they fail to draw a conclusion (as their knowledge expands), they may revise their conclusions. In other words, humans have certain axioms in their mind (knowledge) that they use to prove certain theorems (conclusions). Under non-monotonic reasoning, if a proof fails, we recognize this failure of the proof and predicate some action on this failure (I \textit{don't know} direction to a friend's house, I'll use a GPS system). Classical logic-based methods gets stuck or fails if an automatic proof is not able to proceed forward: all that can be reported is that the proof failed. For example, if we code reachability in graph as two axioms: (i) node {\tt B} is reachable from node {\tt A} if node {\tt A} has a direct edge to node {\tt B}, and (ii) Node {\tt C} is reachable from node A, if there is a direct edge from {\tt A} to {\tt B} and, recursively, {\tt C} is reachable from {\tt B}. If we have a graph that contains an isolated node {\tt U}, then we cannot conclude that node {\tt U} is unreachable from another node {\tt X} inside the graph. Our proof for {\tt reachable(X, U)} will fail. In FOL, \textit{to infer unreachability, we will have to define separate axioms for the concept of unreachability}. In a non-monotonic logic (such as ASP), failure of proof of node {\tt U} being reachable prompt us to conclude that {\tt U} is unreachable through the use of negation-as-failure (NAF). Day to day human-style commonsense reasoning is indeed non-monotonic. When humans design data-structures or write code, they invariably rely on non-monotonic (commonsense) reasoning. Thus, our premise is that if domain experts write declarative specifications in a non-monotonic formalism such as ASP, then efficient imperative code will be much more easily derivable from the specification. We illustrate this idea by automatically deriving efficient imperative code for linked-list and external binary search tree operations.
     %We assume that birds fly, but if we encounter a non-flying bird 
     %(e.g., penguin), we make amends and conclude that normally birds 
     %fly but there are exceptions. 
     \hphantom{the}

     Thus, we generate imperative programs for algebraic operations of pointer data structures such as Linked Lists and External Binary Search Trees using specifications written in ASP. Our work is to be contrasted with Program Verification and other approaches for Program Synthesis that are based on constraint solving \cite{gulwani2008program, srivastava2010program}. Program verification using constraint solving involves reducing an imperative program to a set of constraints and proving that the constraints are satisfiable. This often involves additional logical formalisms to be encoded in the constraint solving system. For example, to verify a linked list program in practice, Separation Logic \cite{reynolds2002separation, cacmseparation} is used which is a formalism on top of Hoare-logic \cite{hoare1969axiomatic}. On the other hand, current state of the art program synthesis generates syntactic programs and verifies them given an Input-Output specification. Our work varies from both these approaches. We assume that a domain expert provides a declarative behavioural specification of a pointer data structure from which an imperative program is synthesized. The behavioural specification contains representation, properties and primitive operations of the data structure.  Representation encodes the domains involved in describing the data structure and its associated well-formedness conditions. Properties encode abstractions used in the respective data structures. Common properties for pointer data structures include reachability of nodes, relationships between nodes present in the data structure and so on. Primitive operations are the permitted read and write operations that can perform operations on only a part of the data structure.  The program generated is correct-by-construction as in the case of Program Synthesis and is performed without the need for additional logics. The resultant program is nothing but a composition of the allowed primitive operations.  
     Our approach presents an inductive proof-technique based on partial deduction \cite{komorowski1992introduction} of logic programs, to generate the imperative programs. Currently, our framework can generate code for \textit{insert} and \textit{delete} operations of Linked List and External BSTs. 
     %GG: stuff removed from here and moved up
     Our work assumes the domain expert encodes the representation and properties of Linked List insert, delete operation in ASP. Once the specification is provided, we evaluate the specification symbolically as part of an inductive proof to look for deterministic imperative program patterns. If the models of the specification (Logic Program) exhibit certain uniform behaviors then the models can be translated to an imperative program. We explain the complete proof-technique for Linked List insert operation end-to-end in this paper: from the description of the specification to the proof that generates the program. Another appeal of our work is that the specification is more semantics driven as it is provided by the domain expert. 
     %GG: commented:
     %There is more information captured by the behavioural specification than an input-output specification which lifts the burden of guessing a large space of program sketches. We next give a background in ASP in the next section followed by an end-to-end demonstration of the synthesis in sections thereafter.  

\section{Background}

\noindent{\bf Answer Set Programming:} Answer Set Programming is very similar to First-Order Logic except that the deductions performed may be non-monotonic. Inferences that were presumed to hold may be retracted in light of new information. Answer Set Programming has both model-theoretic \cite{gelfond1988stable} and a proof-theoretic semantics \cite{arias2018constraint}. A typical Answer Set Program is a collection of rules of the form:
 \begin{enumerate}
 \item
 $p \; \leftarrow \;$
 \item
 $p \; \leftarrow \; q_1,~q_2,~\ldots,~q_m,~not \ r_1,~not \ r_2,~\ldots,~not \ r_n    \{m \geq 0,~n \geq 0\}$ 
 \item
 $\; \leftarrow \; p',~q',~\ldots,~r'.$ 
 \end{enumerate}
 The first rule form depicts raw facts that are unconditionally true. The second rule is the implication used in Prolog ie. $p$ is true if the literals in the body of $p$ are true. For the second rule form, the literals $q_1,~q_2,~\ldots,~q_m$ constitute the positive literals in the body of $p$ and likewise, the literals $r_1,~r_2,~\ldots,~r_n$ denote the negative literals. Negated literals take a special place in ASP. Negated literals prefixed with $not$, stand for negation-as-failure (NAF) as opposed to classical negation in First Order Logic (FOL). A negative literal has a truth value $true$ if the literal has no \textit{proof} from the rules specified in a program (or theory). The third rule form represents constraints on the truth of literals. That is, the conjunction of literals in the constraint, as in rule form (3) cannot be simultaneously true. 
 An answer set or a model $M$ of a normal logic program $\Pi$ is a set of satisfiable literals in the program under the negation-as-failure semantics. An answer set program $\Pi$ may have more than one model which gives rise to the non-monotonicity of its inferences. 
 The set of models $M_{\Pi}$ of $\Pi$ are commonly referred to as Stable models and the corresponding semantics of ASP is called the Stable Model Semantics. An alternative way to interpret rule form (3) or constraints is, the rule disallows the conjunction of literals in the constraint to be part of any model of $M$ of $\Pi$. 
 \par The simplest program with two models is the set of two rules as follows: $\{p \; \leftarrow \; not \ q,~q \; \leftarrow \; not \ p\}$, which has two models, one model containing just the literal $\{p\}$ and the other model containing just $\{q\}$. The aforementioned set is characteristically termed as an \textit{even loop over negation} as both $p,~q$ are mutually defined in a cycle containing 2 negations. Even loops serve as generator of possibilities, and their counterparts: \textit{Odd loops} serve to limit the possibilities. For example, the simple odd loop: $\{p' \; \leftarrow \; q',~not \ p' \}$ precludes $q'$ from ever being true, which imposes a constraint on the truth of $q'$. %GG: parenthetical comment connecting constraint and odd loop added below
 (Note that a constraint is syntactic sugar for an odd loop over negation: $\{p' \; \leftarrow \; q',~not \ p' \}$ is also written as $\leftarrow q'.$)
 This constraint is valid when $p'$ is referenced only in the odd loop, to block $q'$. The odd loop constraints $q'$  because if $\{q',~not \ p'\}$ were a model, $\{p'\}$ would be inferred which is inconsistent. Likewise, if $\{q',~p'\}$ were a model, then this would imply $p'$ is inferred via the odd loop, which means $\{not \ p'\}$ should also be true. Predicate symbols can involve variables ie. one can express rules such as $\{p(X_1,~X_2,~\ldots,~X_k) \; \leftarrow \; q(Y_1,~Y_2,~\ldots),~\ldots,~not \ r(Z_1,~Z_2,~\ldots)\}$. 
 \par In the model-theoretic semantics, the models of an ASP program are found by reducing the program to a procedure involving SAT-solving \cite{gebser2016theory}. For first order logic programs, the programs are assumed to be Datalog programs \cite{gallaire1989logic} with variables ranging over finite domains. The ASP solver then propositionalizes the program by grounding the variables and obtains the Answer Sets. In contrast, the Proof-Theoretic semantics of ASP starts with a query. A query is a conjunction of literals, similar to a query in Prolog. Once a query is issued, all the sub-goals leading up to the query are explored with the appropriate constraint checking. This results in a proof-tree for the query and all literals present in the proof-tree constitute the ``partial" stable model of the given ASP program. Finally, we also need the notion of \textit{defeasible rules}. Due to non-monotonicity of inferences, presence of some literals in an answer set can render some rules in an ASP program infeasible. Considering the simple even loop mentioned previously, the answer set containing just $\{p\}$ renders the rule $q \; \leftarrow \; not \ p$ infeasible. Details of ASP can be found elsewhere \cite{gelfond2014knowledge}.
 
\medskip 
\noindent{\bf Partial Deduction of Logic Programs:}
   Logic programs represent both a first order theory and computational artifacts where a computation begins with an issued query \cite{komorowski1992introduction}. For a rule $p \; \leftarrow \; q$, proposition $p$ is inferred if $q$ is true. This corresponds to standard implication in Logic. Computationally, a query $?- p$ triggers a recursive top-down search of the rules defining $p$ and succeeds if the search terminates at a fact. This is widely known as the SLD-tree expansion of a query (goal) in Logic Programming \cite{lloyd2012foundations}. The SLD-computation and the logical deduction of inferring $p$ from $q$ are equivalent. As an application in  partial evaluation, the top-down search of SLD-trees can be directed to explore only part of the search space while selectively not evaluating another part of the search space. For example, for the rule $\{p \; \leftarrow \; q,~r\}$ and the query $\{?- p\}$, $q$ may have a proof but $r$ may not. Assuming the definition of $r$ may be incomplete, the rule for $p$ may be re-written as $p \; \leftarrow \; r$. The re-written rule is a simplification of the original rule and is referred to as ``residual code". One can view partial evaluation as performing deduction up to known information. Hence this process of partial evaluation is also termed Partial Deduction. 

\medskip 
\noindent{\bf Hoare-Triples in Imperative Programs:}
    The axiomatic semantics of imperative programs is the logical formalism due to Hoare \cite{hoare1969axiomatic}. In this logic, every program   statement has meaning according to the logical conditions of the variables that hold before and after the statement executes. For a program statement $S$, the Hoare-triple written as $\{P\} S \{Q\}$, asserts that if a precondition $P$ holds before the statement $S$ executes, then $Q$ should hold after $S$ terminates. $Q$ is referred as the post-condition of $S$. Two statements, $S_1,~S_2$ in that temporal order, stand for two Hoare-triples, $\{P_1\} S_1 \{Q_1\}$ and $\{P_2\} S_2 \{Q_2\}$ where the post-condition $Q_1$ (of $S_1$) implies the precondition $P_2$ (of $S_2$). Consequently, given two Hoare-triples satisfying the postcondition-precondition implication, they represent two statements, one followed by another. As we show in our proof, the declarative specification can be transformed into a set of temporal facts, which can be interpreted as Hoare-triples. This constitutes the basis of our imperative program synthesis.  

\medskip 
\noindent{\bf Steps in Synthesis:}
     We describe the end-to-end synthesis of linked-list insert operation in this paper. We first describe the specification of linked-list representation in terms of nodes and edges, then describe transitive properties of nodes involved in a list such as reachability. For pointer data structures, their algebraic operations written as an imperative program usually consist of a traversal part followed by a destructive update part. The traversal part visits nodes in the data-structure and identifies the right ``window'' where the destructive update might be performed. The traversal part, at least for sequential data structures, does not visibly modify the data structure. The destructive update part updates the pointers of nodes visited during the traversal part according to the constraints imposed by the algebraic operation and other correctness conditions. For example, a key inserted into a linked-list must preserve the ordering of keys within the list. Thus, we break up the synthesis into two sub-tasks. The first sub-task synthesizes the imperative code for destructive update and also produces a precondition necessary to perform the destructive update. The second sub-task generates the traversal code such that the post-condition of the traversal is the same as the precondition from the first sub-task. Thus, there are two specifications, $\pispec$ for destructive update and $\pitrav$ for traversal. We generate the imperative code $\Pi_1$ and $\Pi_2$ respectively for $\pispec$ and $\pitrav$ and combine them into a single program that performs the desired algebraic operation. Section 3 describes the encoding of lists in ASP followed by complete description of $\pispec$. Section 4 describes a proof by induction required to produce an imperative program for destructive update. Section 5 describes the specification $\pitrav$ for the traversal part and generation of traversal code from $\pitrav$. Finally, in Section 6 the partial programs from Section 4 and 5 are composed into a single program along with formalizing the conditions involved in synthesis. Finally, we conclude with closing remarks and future work.

\section{Specifying Linked Lists as an Answer Set Program}
\subsection{Preliminary Logical Relations}
Let us consider the task of inserting a key into a linked list. The linked-list is assumed to be represented as a set of heap locations connected in a chain by pointers. The representation is encoded as a set of domains\footnote{$X, Y, Z$ denote nodes, $T$ denotes time, $K, k$ represent keys, $\tau$ is the target node to be inserted, \\$\tau$ and $target$ are used interchangeably. Uppercase letters are variables (non-ground terms), lowercase letters are ground terms} and logical relations in ASP. A node (heap-cell) in the linked-list corresponds to the domain \textit{node}, ie. \textit{node(X)} denotes \textit{X} is a heap-cell. Pointers from one node to another in a linked-list correspond to the \textit{edge} relation. \textit{edge(X,~Y)} denotes the \textit{next} pointer of node \textit{X} is points to node \textit{Y}. Every node has an associated key \textit{K}, modelled as \textit{key(X,~K)}. The domain of keys themselves are a numeric domain represented by the relation \textit{num}. An operation we can perform on a list is modifying the pointer linkage. Suppose we want to link node \textit{X} to node \textit{Y}, we represent it using the action \textit{link(X,~Y)}. As the insert operation performs steps that modify the data structure, relations that vary over time have an extra time argument \textit{T}\footnote{We assume discrete time in this paper for simplicity; it is possible to model continuous time using constraints over real numbers.}. 
%GG: footnote above inserted
Therefore, the relation \textit{edge(X,~Y,~T)} means node \textit{X} points to node \textit{Y} at time \textit{T}. Similarly actions executing at time \textit{T} are represented as \textit{link(X,~Y,~T)}. 

\subsection{Effects of Actions}
Our Linked-list specification has only one action, denoted by \textit{link}. The action \textit{link(X,~Y,~T)} unconditionally makes \textit{X} point to \textit{Y} at time \textit{T+1}. This means that \textit{edge(X,~Y,~T+1)} is true when \textit{link(X,~Y,~T)} is true, given by the following rule:
{\small
\begin{align}
     \tag{New-Edge}
     edge(X,~Y,~T+1)  \; \leftarrow \;  link(X,~Y,~T)
\end{align}
} 
Along with effects of actions, one needs to specify that a previous state is retained when no action takes place. This is commonly referred to as the rules of inertia which roughly means, whatever was the state, continues to be the state unless acted upon. For instance, if a node \textit{X} was not modified through pointer-linkage at time \textit{T}, then \textit{X} would retain the same ``next" pointer at  \textit{T+1}. The inertial rules are as follows:
 {\small
 \begin{alignat}{1}
    \tag{Edge-Inertia}
    edge(X,~Y,~T+1)  & \; \leftarrow \; edge(X,~Y,~T),~\mathit{not \ modified(X,~T)} \\
    \tag{Node-Modified}
    \mathit{modified(X,~T)} & \; \leftarrow \; not \ link(X,~Y,~T)  
\end{alignat}
} 

\subsection{Abstractions}
Our specification is thus far incomplete. We need notions of what constitutes a linked-list, what it means for a key to be present in the list and so on. For encoding the list structure, we assume that every linked list has two sentinel nodes called \textit{h} and \textit{t} which are situated at the two ends of the list. Node \textit{h} has the key with the least possible value whereas node \textit{t} has the key with the largest possible value. All other nodes present in the list have keys in ascending order. These conditions are represented by the \textit{admissible} abstraction, standing for the mathematical definition of a well-formed list. First, we have the base case:
{\small
\begin{align}
\tag{Admissible-Base}
admissible(T) & \; \leftarrow \; edge(h,~t,~T),~key(h,~K1),~key(t,~K2),~K1 < K2 \\
\tag{Admissible-Recursive}
\begin{split}
admissible(T) & \; \leftarrow \;   edge(h,~X,~T),~key(h,K1),~key(X,~K2), \\
& \phantom{\; \leftarrow \;}~~ K1 < K2,~\mathit{suffix(X,~T)} 
\end{split} \\
\tag{Suffix}
\begin{split}
\mathit{suffix(X,~T)} & \; \leftarrow \; edge(X,~Y,~T),~key(X,~K1),~key(Y,~K2),~K1 < K2, \\ & \phantom{\; \leftarrow \;}~~ \mathit{suffix(Y,~T)} 
\end{split} \\
\tag{Suffix-t}
\mathit{suffix(t)} & \; \leftarrow \; 
\end{align}
}
\indent \textit{Suffix} specifies that a chain of connected nodes terminate at the \textit{t} node. Trivially, node $t$ terminates at $t$.

In addition to \textit{admissible} we need the notion of reachability of nodes in the linked list. The abstraction {reachable} is the transitive closure of \textit{edge}:
{ \small
\begin{align}
\tag{Reachable-Head}
reachable(h,~T) & \; \leftarrow \;   \\
\tag{Reachable-Recursive}
reachable(X,~T) & \; \leftarrow \;  edge(Y,~X,~T),~reachable(Y,~T)
\end{align}
}
Finally, to encode the notion of keys present in the list we need the \textit{present} relation. A key \textit{K} is present in the list if it is associated with a reachable node.
{\small
\begin{align}
\tag{Key-Present}
present(K,~T) \; \leftarrow \; key(X,~K),~reachable(X,~T).
\end{align}
}
Our objective is to insert a specific target key in the linked list. The target key can be annotated as \textit{target(K)} for some number \textit{K}. Our goal is to have the target key (not present initially in the list) to be present at some time \textit{T}, written as:
{\small
\begin{align}
\tag{Objective}
goal(T) \; \leftarrow \; target(K),~not \ present(K,~0),~present(K,~T).
\end{align}
}
\vspace*{-7mm}

\subsection{Safety Constraints}
  The specification is unsound yet as it supports arbitrary changes to the world, e.g., one possible solution is to execute the action \textit{link(h,~target,~0)} resulting in target key being present at time \textit{T = 1} trivially satisfying the goal. Clearly, this is unacceptable as \textit{admissible(1)} is violated. All safety constraints are easily expressed in ASP. We require the list be well-formed at all times, written as:
 {\small
 \begin{align}
 \tag{Admissible-Universal}
 & \; \leftarrow \;  not \ admissible(T).
 \end{align}
 }
 Currently, both \textit{link(X,~Y,~T)} and \textit{link(X',~Y',~T)} can execute at the same time. To prevent this, we add constraints that disallow the actions to happen simultaneously.  
{\small
\begin{align}
\tag{Link-Single-Node}
& \; \leftarrow \; link(X,~Y,~T),~link(X',~Y',~T),~X \neq X'
\end{align}
}
We should also impose constraints on the \textit{edge} relation. This will avoid redundant actions such as \textit{link(t,~h,~T)} from taking place. Another important restriction on edges is that there cannot be a single node point to two different nodes. Although this is impossible in an imperative program, this is very much possible in a logical theory. If any logical consequence that is undesirable is not constrained, it will take place and show up in the answer set. This would make the specification unsound and hence the synthesized program unsound. 
{\small
\begin{align}
\tag{No-Node-Beyond-Tail}
& \; \leftarrow \;  edge(t,~X,~T) \\
\tag{No-Node-Prior-Head}
& \; \leftarrow \;  edge(X,~h,~T) \\
\tag{No-Self-Loop}
& \; \leftarrow \; edge(X,~X,~T) \\
\tag{No-Multi-Edge}
& \; \leftarrow \; edge(X,~Y,~T),~edge(X,~Z,~T),~Y \neq Z
\end{align}
}
A final unintended consequence is that any key already part of the list should not disappear while inserting the target key. The respective constraint is give below:
{\small
\begin{align}
\tag{No-Key-Loss}
\; \leftarrow \; present(K,~0),~not \ present(K,~T)
\end{align}
}

\subsection{Circular Negation to Generate Actions}
Although we have effects of actions specified, there is no reason for the action to hold at any point of time. There must be a way to choose an arbitrary action at a point in time. This is achieved via even loops in ASP. 

{\small
\begin{align}
 \tag{Generate-Link-Action}
\begin{split}
link(X,~Y,~T) & \; \leftarrow \;  not \ neg\_link(X,~Y,~T) \\
neg\_link(X,~Y,~T) & \; \leftarrow \;  not \  link(X,~Y,~T)
\end{split}
\end{align}
}
Once the specification is complete, the domain expert can test his theory with a set of input facts. The input facts represent the initial state of the world. The input facts are some set of concrete nodes, keys and edge relations. The domain expert must test his theory with various finite instances of input facts. The domain expert then needs to convince himself that the models (answer sets) conform to his understanding of linked lists. We denote the just described specification (ASP Program) as $\pispec$.  For instance, for the following sample inputs the sample output answer sets are shown below.
\begin{mdframed}[leftmargin=10pt,rightmargin=10pt]
$\mathit{Input \ facts}: node(h). \ node(t). \ node(a). \ node(\tau).
             \ key(h,2). \ key(t,5). \ key(\tau, 4) \newline num(2).  \ num(5). \ num(4). \ time(0..2). 
             \ edge(h,~t,~0). \ target(4).$
\end{mdframed}
Here, 4 is the target key, represented by the node $\tau$. A sample answer set is shown below:
\begin{mdframed}[leftmargin=10pt,rightmargin=10pt]
$ Answer \ Set: \ link(\tau,~t,~0). \ link(h,~\tau,~1). \ goal(2).$
\end{mdframed}

If the domain expert does not provide enough time steps for the ASP program, then the output would be simply \textit{unsatisfiable}. That is, there exists no plan that could achieve the goal in the prescribed time steps. \mycomment{Usually, the domain expert starts with time \textit{T = 0} and incrementally tests the specification with higher values of \textit{T}.} In the above example, the least amount of time required to solve the problem is \textit{T = 2}. \mycomment{Because the input facts are general enough, we can be sure that there exists no solution for \textit{T $<$ 2}. However, this needs to proved for all input instances.} Let $\pispec[t]$ denote the grounded program allowing time steps to range from $T = 0 \ to \ t$. \mycomment{Once the specification is deemed correct by the domain expert, we are ready to synthesize the program that works on all inputs which can be infinite.
Since the satisfiability of $\pispec[t]$ is checked for a prescribed maximum time $T = t$, we denote $\pispec[t]$ for the grounded theory allowing time steps to range from $T = 0 \ to \ t$.} 
The program derivation task is discussed next. 

\section{Deriving Linked List Program from ASP Specification}

  \mycomment{So far we have discussed how a domain expert could model a linked list using ASP. Once the expert is confident of the concrete answer sets for several instances of concrete input facts, it is reasonable to expect that the ASP program might exhibit a deterministic imperative program structure which could handle all sets of input facts. Because every list can be distinguished from another by the set of keys it holds, it is a difficult task to derive a program by analyzing lists purely based on sets of keys they hold.}
  We can leverage the structural nature of lists to derive a deterministic program from $\pispec$.  Every admissible list has a well-defined structure and one can perform an inductive argument on the structure of the list without being concerned about the concrete sets of keys an arbitrary list may carry. Thus, the inductive nature of the data structure (pointer-based or otherwise) is crucial to transform the ASP specification into a deterministic imperative program. 
  \mycomment{Our task is shown diagrammatically in Figure 1. The meaning of ``Extended Linked List ASP Program" will be explained when discussing the evaluation of symbolic inputs.
  \begin{figure}
      \centering
      \includegraphics{fig1.PNG}
      \caption{Using ASP Program to derive imperative programs}
      \label{fig:my_label}
  \end{figure}
  } 
  \subsection{Symbolic Inputs/Outputs via Inductive Definitions} 
  \mycomment{To understand where symbolic inputs come from, let us look at the definition of \textit{admissible} from the previous section.}
  Symbolic inputs are obtained using the definition of \textit{admissible}. The predicate
  \textit{admissible}, has a base case which defines a valid empty list, and an inductive case which defines lists of increasing size. This exactly depends on how many times \textit{suffix} rule was applied for \textit{admissible} to hold true. If \textit{suffix} was not applied at all, then we have the base case. If \textit{suffix} was applied once, it corresponds to a linked list of size at least one. If \textit{suffix} was applied twice, it corresponds to a linked list of at
  least size two and so on. The base cases and the inductive cases together constitute the symbolic inputs.  To perform synthesis, we first run the specification with the base case, and then run the specification with the inductive case. If the ASP program is satisfiable with both the base case and the inductive case of \textit{admissible(0)}, then the ASP program is satisfiable for all linked lists. Correspondingly, the answer sets (symbolic) represent an imperative program. The base case and inductive case are shown below:
\begin{mdframed}[leftmargin=10pt,rightmargin=10pt]  
   \texttt{Base case:} 
   $admissible(0) \; \leftarrow \;  edge(h,~t,~0),~key(h,~k_h),~key(t,~k_t),~k_h < k_t$
\end{mdframed}
   If there exists an answer set with the base case definition of admissible(0),~then we check for an answer set with the inductive case.
\begin{mdframed}[leftmargin=10pt,rightmargin=10pt]    
   \texttt{Inductive cases:} 
 \begin{align*}
   admissible(0) &  \; \leftarrow \; edge(h,~x,~0),~\mathit{suffix}(x,0),~key(h,~k_h),~key(x,k_x),~k_h < k_x \\ 
   admissible(0) & \; \leftarrow \; edge(h,~x,~0),~edge(x,~y,~0),~\mathit{suffix}(y,~0),~key(h,~k_h), \\ 
            & \hphantom{\; \leftarrow \;}~~ key(x,k_x),~key(y,~k_y),~k_h < k_x,~k_x < k_y
\end{align*}
\end{mdframed}
   If there exists an answer set in the above inductive cases, then our proof that there exists a symbolic answer set for all linked lists would be complete. Determining the existence of answer set in our proof method corresponds to evaluating the ASP Program by symbolic means for the base case and the inductive cases. Notice that the definitions shown above do not have concrete values for keys. These symbols will be interpreted in a suitable way in the following section to prove satisfiability of the ASP specification. 
  
  \subsection{Abstracting Concrete Symbols}
    Every expression in the ASP program for linked-list is built up of variables from the domains of Nodes, Keys and Time. We explain how they are each handled in turn. 
    \subsubsection{Handling Time}
      \ Time is always handled concretely. For every concrete time step starting from $T = 0$ onwards, we systematically check for satisfiability. To check satisfiability for time $T = t$, we check whether \textit{goal(t)} is satisfiable. If not, we proceed to check for satisfiability of \textit{goal(t+1)} and so on. Because time is treated concretely, we simply rely on the ASP grounder to check constraints that are dependent on time.
    \subsubsection{Handling Nodes}
      \  Nodes are treated symbolically. This is because, node symbols generated by the definition of \textit{admissible(0)}, are used to prove (as opposed to concrete evaluation) the satisfiability (or unsatisfiability) of the ASP Program. As mentioned before, \textit{admissible(0)} symbolically specifies the initial state of the list. To check satisfiability of the ASP program, we also need to assert that the final state (\textit{goal(T)}) of the linked list is true. In the case of linked list insert operation, the final state asserts the reachability of the node that has the target key. In addition to reachability, the final state also represents a well-formed list. This implies that the list in the final state should take the form specified by \textit{admissible}. That is, the list should be of the form:
       \textit{edge(h,~x,~t'),~edge(x,~x1,~t'),~\ldots,~edge(z,~y,~t'),~edge(y,~t,~t')} at some time $T = t'$. We know that in order for some node \textit{target\_node} carrying the target key to be reachable, one of the nodes \textit{\{x,~x1,~\ldots,~z,~y\}} must be equal to the target node $\tau$. Hence we need a notion of equality of nodes. This is specified by the logical relation \textit{eq\_node(X,~Y)} meaning node \textit{X} equals node \textit{Y}. 
     \subsubsection{Handling Keys}
       \  Keys that nodes carry are also treated symbolically. A key \textit{k1} can be less than a key \textit{k2} in the definition of \textit{admissible}. This necessitates the requirement of the arithmetic inequality relation, \textit{lt(K1,~K2)} denoting key \textit{K1} is less than the key \textit{K2}. Further, equal nodes have equal keys. Although this is obvious to a domain expert, it has to be encoded via the definition of the \textit{eq\_key(K1,~K2)} relation which denotes that key \textit{K1} is equal to key \textit{K2}.
    We have described notions of equality of nodes, keys and arithmetic inequality. These relations abstract concrete expressions in a concrete evaluation. 
    \mycomment{They abstract the concrete expressions and are reasoned upon by symbolic means, a technique commonly referred as Predicate Abstraction \cite{ball2001automatic}.}
    To perform abstract execution of the ASP Program, it must be extended with rules using these abstractions (logical relations) in a consistent manner. In doing so, few of the rules in the original program need modifications to support the abstractions. For instance, every occurrence of the expression $X < Y$ should be replaced by $lt(X,~Y)$. The complete extension of the ASP program is discussed in the following section.
  \mycomment{
  \subsection{Intuitive Argument for Program Derivation}
   Our task is to find a sequence of actions that transition the initial state of the linked list to a desired final state (goal) of the list. The goal in our case is inserting a key into a linked list as specified by \textit{goal(T)} in the previous section. We systematically test for satisfiability starting from time $T = 0$. If the ASP program is not satisfiable within $T = 0$,~then we consider satisfiability for $T = 1$ and go on until we find a solution. Let us assume that the program is satisfiable for some time $T = t$. To derive the program, we start with the assumption that at time $t$, $goal(t)$ holds, subject to all the constraints in the specification. One such constraint is \textit{admissible}. That is, \textit{admissible(t)} holds for all time steps starting from time $T = 0$. Because \textit{admissible(t)} describes the complete state of the list for every time step $0 to t$, we generate symbolic nodes, edges, and keys allowed by the definition of \textit{admissible(0)}. 
   }
  \subsection{Extending Linked List Specification to Symbolic Form}
 We introduced the logical relations that enable their symbolic interpretation in the previous section. The new relations introduced, \textit{eq\_node,~eq\_key} and  \textit{lt} must be used in a consistent manner in the original program. This implies that certain rules in the original program which use the concrete notions of node equality (=), node inequality (!=), arithmetic inequality ($<$), key equality (=) are replaced with their logical relation counterparts. The following table  provides the mapping between the concrete operators and logical relations.
       \vspace{-0.8cm}
       \begin{table}
    \scriptsize
    \centering
    \begin{tabular}{|c|c|c|}
        \hline
        Relation name & Concrete Symbol  & Abstract Predicate \\
        \hline
        Node equality &   =             &  eq\_node  \\
        Node inequality & !=            & not eq\_node \\
        Key equality & =               & eq\_key \\
        Key inequality & !=         & not eq\_key  \\
        Arithmetic inequality & $<$   & lt \\
        \hline
    \end{tabular}
    \end{table}
    \vspace{-0.7cm}
    Because the symbolic execution is part of mathematical proof to check satisfiability, we can naturally make use of properties of arithmetic inequality and equality relation in general. These properties are not implicit in the ASP Program. Because the ASP solver is not a Theorem-prover, the properties have to be encoded explicitly. The transitivity of equality, arithmetic inequality is captured explicitly. In addition, equality is symmetric. The rules for reflexive, symmetric and transitive relations are provided below:
       \begin{alignat}{1}
          \tag{Eq-Node-Transitive}
          & eq\_node(X, Z) \; \leftarrow \; eq\_node(X, Y), eq\_node(Y, Z) \\
          \tag{Eq-Key-Transitive} 
         &  eq\_key(K1, K3) \; \leftarrow \;  eq\_key(K1, K2), eq\_key(K2, K3) \\
           \tag{Arith-Ineq-Transitive}
         &  lt(K1, K3) \; \leftarrow \;  lt(K1, K2), lt(K1, K3) \\
           \tag{Eq-Node-Commutative}
         &  eq\_node(X, Y) \; \leftarrow \;  eq\_node(Y, X) \\
           \tag{Eq-Key-Commutative}
         & eq\_key(K1, K2) \; \leftarrow \; num(K1), num(K2), eq\_key(K2, K1) \\
         \tag{Trichotomy-1}
         & eq\_key(K1, K2) \; \leftarrow \; not \ lt(K1, K2), not \ not\_lt(K1, K2) \\
         \tag{Trichotomy-2}
         & lt(K1, K2) \; \leftarrow \; not \ eq\_key(K1, K2), not \ not\_lt(K1, K2) \\
         \tag{Trichotomy-3}
         & not\_lt(K1, K2) \; \leftarrow \; not \ lt(K1, K2), not \ eq\_key(K1, K2) \\
         \tag{Circular-Neg-Eq-Node}
         & eq\_node(X, Y) \; \leftarrow \; not \ neg\_eq\_node(X, Y) \\
         \notag
         & neg\_eq\_node(X, Y) \; \leftarrow \; not \ eq\_node(X, Y) \\
         \tag{Not-Lt-Inverse}
            & not\_lt(K1, K2) \; \leftarrow \; lt(K2, K1) \\
         \tag{Neq-Key-Inverse-1}
         & not\_eq\_key(K1, K2) \; \leftarrow \; lt(K2, K1) \\
         \tag{Lt-Inverse}
         & lt(K1, K2) \; \leftarrow not\_lt(K2, K1) \\
         \tag{Neq-Key-Inverse-2}
         & not\_eq\_key(K1, K2) \; \leftarrow not\_lt(K1, K2) \\
         \tag{Circ-Neg-Eq-Key}
         & eq\_key(K1, K2) \; \leftarrow not \ neg\_eq\_key(K1, K2) \\
         \notag
         & neg\_eq\_key(K1, K2) \; \leftarrow not \ eq\_key(K1, K2)
   \end{alignat}
    The extended Linked list specification is by no means complete. \mycomment{Domains such as nodes, keys must be interpreted in their symbolic form while considering equality and inequality of nodes and keys.} Rules or constraints that use equality or inequality or arithmetic inequality must be rewritten with predicates \textit{lt,~eq\_node,} and \textit{~eq\_key}.  We denote the extended ASP program by $\pisym$. Similar to $\pispec[t]$ the finite grounding of the program with prescribed maximum time $T = t$ is denoted by $\pisym[t]$. The advantage of using $\pisym$ is that we can feed $\pisym[t]$ to the same ASP solver. That is, symbolic evaluation is reduced to concrete evaluation. We can generate symbolic answer sets and reason over them in the same way as concrete answer sets. Another important point to note is that during partial deduction, evaluation of predicates and constraints on parts of the list may be suspended. This gives rise to the new domain of nodes modelled by the \textit{suspended} relation. For soundness of the partial deduction, suspended nodes should be left untouched and not be modified. The rewrites that include suspended nodes are given below: 
    
       \begin{align}
       \tag{Rewrite-No-Self-Loop}
       & \; \leftarrow \; \textit{\textbf{not  \ suspended(X)}}, edge(X, Y, T), \textit{\textbf{eq\_node(X, Y)}} \\
       \tag{Rewrite-Admissible-Base}
      &  admissible(T) \; \leftarrow \; edge(h,~t,~T), key(h,~K_1), key(t,~K_2), \boldsymbol{lt(K_1,~K_2)}   \\
       \tag{Rewrite-Admissible-Recursive}
     & admissible(T) \; \leftarrow \; edge(h,~X,~T), key(h, K_1), key(X, K_2), \boldsymbol{lt(K_1,K_2)}, \mathit{suffix(X,~T)} \\
    \tag{Rewrite-Suffix}
       & \mathit{suffix(X,~T)}  \; \leftarrow \; edge(X,~Y,~T), key(X,K_1),key(Y,K_2), \boldsymbol{lt(K_1, K_2)}, \mathit{suffix(Y,~T)}  \\
       \tag{Rewrite-No-Multi-Edge}
       &  \; \leftarrow \;  \textit{\textbf{not suspended(X)}}, edge(X, Y, T), edge(X, Z, T), \textit{\textbf{not eq\_node(Y, Z)}}, \\
        \tag{Rewrite-Link-Single-Node}
        & \; \leftarrow \; \textit{\textbf{not suspended(X)}}, link(X, Y, T), link(X, Y_1, T),  \boldsymbol{not \ eq\_node(Y, Y_1)} \\ 
        \tag{Suspended-Unmodified}\footnote{     As mentioned before, suspended nodes should not be modified.}
        & \; \leftarrow \; \textit{\textbf{suspended(X)}}, link(X, Y, T)
    \end{align}

     Abstractions must be extended due to introduction of equality. Abstractions over equal objects must retain the same truth values. That is, if node $x$ is reachable then so should  every node equal to $x$. The corresponding rules are given below:
    \begin{align}
     \tag{Suff-Ext}
     & \mathit{suffix(X,~T)} \; \leftarrow \; eq\_node(X,~Y), \mathit{suffix(Y,~T)} \\
     \tag{Node-Reachable-Ext}
     & reachable(X,~T) \; \leftarrow \; eq\_node(X,~Y), reachable(Y,~T) \\
     \tag{Key-Present-Ext}
     & present(K,T) \; \leftarrow \; eq\_key(K,K1), present(K1,T) \\
     \tag{Edge-Ext-1}
     & edge(X,~Y,~T) \; \leftarrow \; eq\_node(Y,Z), edge(X,Z,T) \\
     \tag{Edge-Ext-2}
     & edge(X,~Y,~T) \; \leftarrow \; eq\_node(X, Z), edge(Z,Y,~T) \\
     \tag{Eq-Key-Ext}
     & eq\_key(K1,K2) \; \leftarrow \; eq\_node(X, Y), key(X, K1), key(Y, K2) \\
     \tag{Suspended-Eq}
     & \textit{\textbf{suspended(X)}} \; \leftarrow \; eq\_node(X, Y), suspended(Y)
    \end{align}

\subsection{Verifying Satisfiability of $\pisym[t]$ Inductively} 
     Our proof is by induction on length of the lists. Rules \textit{Admissible-Base} and \textit{Admissible-Recursive} define lists of size 0, 1, 2, \ldots and so on. Let us predicate the satisfiability and unsatisfiability of $\pisym[t]$ by $Sat(\pisym[t,~k])$ and $Unsat(\pisym[t,~k])$, respectively, where $k$ represents the length of the list. 
     To check satisfiability of $\pisym[t]$, we need to prove the following:
     \begin{enumerate}[nosep]
         \item  $Sat(\pisym[t,~0])$ is true  (We skip this as it is trivial and lack of space)
         \item  $Sat(\pisym[t,~k]) \Rightarrow Sat(\pisym[t,~k + 1])$ is true
     \end{enumerate}

     \subsubsection{Proof for inductive case}
     \ To prove the inductive case, one should assume lists of length $k$ and reason about them. To do that, we \textit{suspend} the evaluation of \textit{suffix}. The technique employed here is partial deduction. An arbitrary list of length $k$ can be realized with the inductive cases mentioned in Section 4.1.
     It is not inconsistent to assume that \textit{suffix(x,~0)} generates a suffix of length $k-1$. We prove this in \textit{Lemma 3}.
     To complete the proof, we have to show satisfiability for a list of length $k+1$. This can be achieved by using the second inductive definition in Section 4.1.
     By the inductive hypothesis, the target key can be placed between the first two nodes for a list of length $k$. To complete the proof, assume for the sake of argument that the target key cannot be placed between the first two nodes of the list of length $k+1$. That is, we reject a model of the form: $\{edge(h,~X,~t),~edge(X,~x,~t),~eq\_key(k\textsubscript{X},~\tau)\}$. With this additional constraint, we can now check for satisfiability of $\pisym[t]$. This is shown below:
     \begin{mdframed}[leftmargin=10pt,rightmargin=7pt]
     {\small
     \textit{Symbolic Input facts:} \newline 
     \phantom{aaa}$node(h). \ node(x). \ node(y). \ node(\tau).  \ num(k_h). \ num(k_y). \ num(k_x). \ num(k_t). \ num(k_{\tau}). \newline \phantom{aaa} key(h,~k_h). \ key(x,~k_x). \ key(y,~k_y).  \ key(\tau,~k_{\tau}). \ lt(k_h,~k_x). \ lt(k_x,~k_y)$. \newline 
     \textit{New  Constraint:} $\{\; \leftarrow \;  edge(h,~X,~T),~edge(X,~x,~T),~key(X,~K_X),~eq\_key(K_X,~k_{\tau})\}$ 
     \newline \textit{Answer  Set:}
     \newline \phantom{aaa}$link(\tau,~y,~0).  \ link(x,~\tau,~1). \ lt(k_x,~k_{\tau}).  \ lt(k_{\tau},~k_y).
     edge(\tau,~y,~2).~edge(x,~\tau,~2).$
     }
     \end{mdframed}
     %%s
     Since $Sat(\pisym[t,~k+1])$ is true, we have checked that $\pispec[t]$ has an answer set for all values of $k$.  
     Similarly, we can perform a proof for $Unsat(\pisym[t,~k])$. In this case, we check for unsatisfiability for a fixed time $T = t$. 
     \subsection{Extracting Program from Symbolic Answer Sets}
      We have the following answer sets for the base case and the inductive case:
      \begin{mdframed}[leftmargin=0pt,rightmargin=0pt]
      {\small
       \textit{Answer  Set  for  base  case:} \newline
       $\phantom{aaa}link(\tau,~t,~0). \ link(h,~\tau,~1). \  key(h,~k_h). \ key(t,~k_t). \ key(\tau,~k_{\tau}). \ 
        lt(k_h,~k_{\tau}). \ lt(k_{\tau},~k_t).$ \newline
       \textit{Answer  Set  for inductive  case:} 
       \newline \phantom{aaa}$link(\tau,~y,~0). \ link(x,~\tau,~1). \  
       key(x,~k_x). \ key(y,~k_y). \ key(\tau,~k_{\tau}). \ 
       lt(k_x,~k_{\tau}). \ lt(k_{\tau},~k_y).$
      }
     \end{mdframed}
      It is easy to see that the actions in both the cases are isomorphic.   We can substitute the terms in the actions with ``Imperative Program Variables" and view them as an imperative program. We prove this in \textit{Lemma 3}. The computations performed by atomic steps in an imperative program are explained logically by the sequence of valid Hoare-triples corresponding to the respective steps. Similarly, the answer sets can be interpreted through the lens of Hoare-triples where the actions supporting the goal represent program steps and the literals before and after the actions constitute the Hoare-triples. More precisely, the conjunction of literals before an action represents its precondition and the conjunction
      %% the conjunction of literals 
      immediately after an action represents its post-condition. Let $Pre(X,~Y,~T)$ denote the precondition involving nodes $X,~Y$ at time $T$ defined as:
      {\small
      \begin{align}
          \tag*{}
          Pre(X,~Y,~T) & \; \equiv \;  edge(X,~Y,~T),~reachable(X,~T),~reachable(Y,~T),~\mathit{suffix(X,~T)},~\mathit{suffix(Y,~T)},
          \\ \notag &  \phantom{\; \equiv \;}~~ key(X,~K_X),~key(Y,~K_Y),~key(\tau,~K_{\tau}),~lt(K_X,~K_{\tau}),~lt(K_{\tau},~K_Y)
      \end{align}
      }
       Then, the imperative program along with Hoare-triples is given below: 
       \begin{mdframed}[leftmargin=4pt,rightmargin=0pt,skipbelow=-10pt]
       {\scriptsize
       \[
       \begin{array}{@{\!\!\!\!\!}rcl}
       \{Pre(x,~y,~0)\} & \mathbf{link(\tau,~y,~0)} & \{Pre(x,~y,~1) \cup \{edge(\tau,~y,~1)\}   
       %% Intermediate 
       \} \; \equiv \; Q \\
       %% $\{Intermediate\} 
       \{ Q \} & \textbf{link(x,~$\tau$,~1)} & \{Pre(x,~y,~2) \cup \{edge(x,~\tau,~2),~edge(\tau,~y,~2),~goal(2)\} \setminus \{edge(x,~y,~2)\}\}
       \end{array}
       \]
       }
       \end{mdframed}

     The reader should take note that the program synthesized is only a fragment of the code for complete insert operation in practice. The fragment generated above is the destructive update of pointers resulting in a new node being added to the list. Before performing the destructive update, the nodes satisfying the preconditions for the first step must be satisfied.  This is achieved by traversing the nodes in the list, one link at a time, to reach the window of the target key. 
     %Further, the conjunction of literals in each of the triples are not minimal. For instance, it is easy to see that \texttt{reachable(y, 0)} is redundant as it is implied by \texttt{reachable(x, 0)} and \texttt{edge(x, y, 0)}. We discuss minimization of literals in the Hoare-triples in a later section.  

\section{Synthesizing Code for Data-structure Traversal}
           Till now, we have shown a proof technique to synthesize the destructive update code $\Pi_{\sigma_{imp}}$for some algebraic operation $\sigma$ of a pointer data-structure.
           In doing so, we also arrived at the precondition $Pre$ necessary to perform the update. 
           As mentioned before, the fragment $\{Pre\}$ $\Pi_{\sigma_{imp}}$ $\{Post\}$ is only a partial program, in the practical sense. We need to perform a traversal of nodes and edges of the data-structure such that $Pre$ is satisfied. The precondition $Pre$ is a conjunction of literals of describing some relationships that the nodes should satisfy in order to perform $\sigma$. 
           In the case of insert operation for a linked list the precondition synthesized is: 
          {\small
          \begin{align}
          \tag*{}
          %% \begin{split}
          Pre(X,~Y) & \; \equiv \; 
                 admissible,~edge(X,~Y),~reachable(Y),~reachable(Y),~\mathit{suffix(X)}, \\ 
                 \notag & \phantom{\; \equiv \;}~~ \mathit{suffix(X)},~eq\_key(target,~k_z),~lt(k_X,~k_z),~lt(k_z,~k_Y) 
          %% \end{split}
         \end{align}
         }
       In the above precondition, the time argument $T$ is removed. This is because of the assumption that the data-structure does not change during traversal ie. traversal is a read-only operation. The variables $X,~Y$ represent two locations in memory that satisfy $Pre$. Now we turn our attention to how a domain expert might specify traversal as an Answer Set Program. Let us denote such a program as $\pitrav$. 
       $\pitrav$ would have the same notions such as \textit{edge, admissible, suffix, reachable} as before but with the time argument $T$ removed. Then, the domain expert would specify a $traversal$ relation capturing the precondition $Pre$. For linked list, the traversal relation is a recursive definition as follows:
      {\small
      \begin{align}
         \tag*{}
         traverse(X,~Y) & \; \leftarrow \; edge(X,~Y),~reachable(X),~reachable(Y),~\mathit{suffix(X)},~\mathit{suffix(Y)}, \\
                         \notag  & \phantom{\; \leftarrow \;}~~           key(target,K),~key(X,~K_1),~key(Y,~K_2),~K_1 < K,~K < K_2
      \end{align}
      }
       The above definition of $traverse(X,~Y)$ is same as $Pre(X,~Y)$ which serves as the base case definition of $traverse$. The recursive case allows for the traversal to `reach' right set of nodes that satisfy $Pre$. 
       {\small
       \begin{align}
       \tag*{}
         traverse(X,~Y) & \; \leftarrow \; 
            edge(X,~Y),~reachable(X),~reachable(Y),~\mathit{suffix(X)},~\\ \notag
            & \phantom{\; \leftarrow \;}~~ \mathit{suffix(Y)},~edge(Y, ~Z),~traverse(Y,~Z)
       \end{align}
       }
       Intuitively, the above definition of traversal seems to be correct. To be certain, we need to verify that $\pitrav$ works in all instances of the data structure. To achieve that goal, we transform the program $\pitrav$ into symbolic form as before. Let this transformed symbolic program denote $\pisymtrav$. We verify that $\pisymtrav$ is satisfiable in all instances inductively, for the base case, and the inductive case. Assuming that $\pisymtrav$ is satisfiable (which it is), we show how the recursive code for traversal can be synthesized. In the above definition $traverse$, it is easy to see that satisfiability of base case of $traverse$ entails the satisfiability of $Pre$. This is because $Pre$ is directly referenced in $traverse$. If not, $Pre$ can be added as a conjunct in the base case definition of $traverse$. 
       
       \subsection{Generating the Recursive Code for Traversal from $\pisymtrav$}
        
        As shown above, $traverse$ is a recursive relation. We therefore generate the recursive procedure satisfying $traverse$. Unlike the synthesis of destructive update, we use a different approach to extract the program form the answer sets.  We assume that there is a well-defined starting point for the traversal provided by the domain expert. The starting point is $\{start\_traversal \; \leftarrow \; traverse(h,~X),~edge(h,~X)\}$. Assuming the computation for traversal begins at $start\_traversal$, we check if $start\_traversal$ is present in every answer set of $\pisymtrav$. If that is the case, then the traversal relation always finds the node \textit{X, Y} satisfying $Pre$. 
        \subsubsection*{Optimizing Traversal Relation} 
        \ We observe that $traverse$ has abstractions such as $reachable,~admissible$ in its definition and cannot be used as is to generate recursive function. A recursive program implementing $traverse$ should use only primitives and not be cognizant of abstractions. Since we have verified that $start\_traversal$ is true in every satisfiable run of $\pisymtrav$, $reachable$ and $admissible$ referenced in $traverse$ can be eliminated. In other words, $traverse$ can be refined to a residual form containing only $edge$ relation, while beginning from $start\_traversal$.  This optimization can be achieved by reasoning over \textit{proof trees} of \textit{start\_traversal}. \textit{Proof trees} are associated with a query. In our case, the query is \textit{start\_traversal}.  Proof trees of a query constitute all the rules (of some program) that were applied in order to satisfy the query. 
      In order to have a proof tree, one needs a Proof-theoretic semantics for ASP. The Proof-theoretic semantics is provided by the s(CASP) system \cite{arias2018constraint}. The s(CASP) system performs a goal-directed execution of a query in an answer set program. s(CASP) also shows the associated proof tree. Therefore, to simplify $traverse$, we obtain the proof trees of \textit{start\_traversal} for two cases (base, inductive) from $\pisymtrav$. A sample run of $\pisymtrav$ with proof tree is shown below:
             Let program $\pisymtrav$ have the following symbolic input facts corresponding to the inductive case. The facts represents the inductive case because the proof of \textit{admissible(b)} is assumed and not fully evaluated. 
     
     \begin{mdframed}[leftmargin=10pt,rightmargin=10pt]
      $Input \ \textit{facts}: 
                \ node(h). \ node(a). \ node(b). \ node(t). \ edge(h, a). \ edge(a, b). \newline admissible(b). 
                \ num(k_h). \ num(k_a). \ num(k_b). \ num(k_t). \ key(h, k_h). \ key(a, k_a). \newline key(b, kb). \ key(t, kt).
                 \ lt(k_h, k_a). \ lt(k_a, k_b). \ lt(k_b, k_t). 
                \  node(target). \newline key(target, k_{ktarget}). \ not\_eq\_key(k_h, k_{target}). \ not\_eq\_key(k_a, k_{target}). \newline 
                 not\_eq\_key(k_b, k_{target}). \  not\_eq\_key(k_t, k_{target}). \
                 \ not\_eq\_node(h, t). \newline not\_eq\_node(h, a).
                 \ not\_eq\_node(a, b). \ not\_eq\_node(b, t).
                 \ not\_eq\_node(h, b).  \newline not\_eq\_node(a, t).
       $             
     \end{mdframed}
      \newpage
      {
         \begin{lstlisting}[basicstyle=\linespread{0.7}\fontsize{7}{13}\selectfont\ttfamily,mathescape]
BEGIN JUSTIFICATION
traverse(h,X) $\rightarrow$ Expand, Unifying with rule head yields traverse(h,X) $\tikzmark{trav-1-start}$
    node(h) $\rightarrow$ Unifying with rule head yields node(h)
    node(Var2) $\rightarrow$ Unifying with rule head yields node(a)
    reachable(h) $\rightarrow$ Unifying with rule head yields reachable(h)
    reachable(a) $\rightarrow$ Unifying with rule head yields reachable(a)
    admissible(h) $\rightarrow$ Unifying with rule head yields admissible(h)
    edge(h, a) $\rightarrow$ Unifying with rule head yields edge(h, a)
    node(Var3) $\rightarrow$ Unifying with rule head yields node(b)
    edge(a,b) $\rightarrow$ Unifying with rule head yields edge(a,b)
    traverse(a,b) $\rightarrow$ Unifying with rule head yields traverse(a,b) $\tikzmark{trav-1-end} \ \ \ \ \ \tikzmark{trav-2-start}$
      node(a)                    
        Coinductive success yields node(a)
      node(b)
        Coinductive success yields node(b)
      reachable(a)
        Coinductive success yields reachable(a)
      reachable(b) $\rightarrow$ Unifying with rule head yields reachable(b)
      admissible(a) $\rightarrow$ Unifying with rule head yields admissible(a)
      edge(a,b)
        Coinductive success yields edge(a,b)
      key(a,Var8) $\rightarrow$ Unifying with rule head yields key(a,ka)
      key(b,Var9) $\rightarrow$ Unifying with rule head yields key(b,kb)
      key(target,Var10) $\rightarrow$ Unifying with rule head yields key(target,ktarget)
      lt(ka,ktarget) $\rightarrow$ Unifying with rule head yields lt(ka,ktarget)
      lt(ktarget,kb) $\rightarrow$ Unifying with rule head yields lt(ktarget,kb)
      num(ka) $\rightarrow$ Unifying with rule head yields num(ka)
      num(kb) $\rightarrow$ Unifying with rule head yields num(kb)
      num(ktarget) $\rightarrow$  Unifying with rule head yields num(ktarget) $\tikzmark{trav-2-end}$
  _nmr_check $\rightarrow$ Unifying with rule head yields _nmr_check
END JUSTIFICATION
        \end{lstlisting}

\AddNoteRight{trav-1-start}{trav-1-end}{trav-1-start}{Recursive case}
\AddNoteRight{trav-2-start}{trav-2-end}{trav-2-start}{Base case}

}

        If $start\_traversal$ is satisfiable, then it is clear that the abstractions \textit{reachable, admissible} would always be satisfied in a proof tree for \textit{start\_traversal}. 
         The residual definition of $traverse$ is given below:
        {\small
        \begin{alignat*}{1}
          \notag
          traverse(X,~Y)  \; \leftarrow \; edge(X,~Y),~key(\tau,~K),~key(X,~K_1),~key(Y,~K_2),~K_1 < K, K < K_2  \\
          \notag
           ~traverse(X,~Y) \; \leftarrow \; edge(X,~Y),~edge(Y,~Z),~traverse(Y,~Z) \hphantom{thequickbrownfoxjump}
         \end{alignat*}
         }
        Notice that the simplified definition is correct with respect to the starting point $\{start\_traversal \; \leftarrow \; traverse(h,~X),~edge(h,~X)\}$. From this residual definition, it is straightforward to generate the recursive function in an imperative language. We mention imperative language in this context because most modern imperative languages support recursion. We assume the artificial syntax of $x.next$ to denote a the ``next" pointer of node $x$. Logically, this is same as $edge(X,~Y)$ for some node $Y$. The synthesized recursive function is shown in the next section.
        
    \section{Combining Destructive Update with Traversal} 
       From sections 4, 5 we have synthesized the partial programs satisfying the destructive update needed to insert a key into a linked list and how to traverse to the right ``window" of insertion respectively. Let the partial programs be represented as $\Pi_1$ and $\Pi_2$ respectively. The precondition of $\Pi_1$ is $Pre(X,~Y)$ and the post-condition of $\Pi_2$ is $Pre(X,~Y)$. Therefore, under Hoare-logic, these two programs compose via the rule of consequence. That is the program $\Pi_{insert}$ = $\Pi_2 \circ \Pi_1$ inserts the target key correctly into a given input linked list. The combined program 
       %GG: inserted
       that is \textit{automatically generated}\footnote{Currently, we compute the generic model for destructive update and simply translate the recursive traversal definition into a recursive function. The whole procedure can be automated end-to-end as shown in the Synthesis Procedure in Section 7} 
       in one of modern representative imperative languages looks as follows:
       \begin{multicols}{2}
       \fontsize{8}{9}\selectfont
       \begin{algorithmic}[0]
       \Function{insert}{$target$}
       \State $(x,~y) \gets$ traverse($target$)
       \State $target.next \gets y$
       \State $x.next \gets target$
       \EndFunction
       \end{algorithmic}
       
       \begin{algorithmic}[0]
       \Function{traverse}{$target$}
       \State \Return traverse($h$, $h.next$, $target$)
       \EndFunction
       \end{algorithmic}
       
       \begin{algorithmic}[0]
       \Function{traverse}{$x$, $y$, $target$}
       \If{$\left(\text{\parbox[c]{0.6\linewidth}{($x.key < target.key$) \textbf{and} \\ ($target.key < y$) \textbf{and} \\ ($x.next = y$)}}\right)$}
          \State \Return $(x,~y)$
       \EndIf
       \State \Return traverse($y$, $y.next$, $target$)
       \EndFunction
       \end{algorithmic}
       \end{multicols}
       %%

%\ssection{Formalizing Program Extraction for Destructive Update}
     \subsection{Class of Data Structures Assumed}
        \medskip\noindent{\textit{Definition}} Heap $H$ is a set of nodes, $L$ is a universe of labels. 
           $E$ is a set of directed edges between nodes, labelled by elements from $L$.
           More precisely, the set of edges $E$ is a relation over $H \times L \times H$.
          No node can have an more than one edge with the same label to another node. That is, 
          \begin{align}
            \tag*{}
             \forall n_1, \forall n_2, \forall n_2',  \forall \ell:~(n_1,~\ell,~n_2) \ \in \  E \ \land \ (n_1,~\ell,~n_2') \ \in\  E \quad \Rightarrow \quad n_2 = n_2'
         \end{align} 
       \textit{Definition} 
          Let $root$ denote a distinguished node in the Heap. 
          A pointer data structure is a two-place relation $D \subset \{root\} \times \mathscr{P}(E)$.
          The relation can be arbitrary. If the relation is $admissible$, then $D$ represents a linked list.  \\
        \textit{Definition} A primitive step $s: H \times H \times L \times \mathscr{P}(E)  \rightarrow \mathscr{P}(E)$ is a function that links two nodes using a label. Its definition is as follows:
         $s(x,~y,~\ell,~e) = (e \setminus \{(x,~\ell,~y') \in e\}) \cup \{(x,~\ell,~y)\}$ \\
         \textit{Definition} An algebraic operation $\sigma_D: H \times \{root\} \times \mathscr{P}(E) \rightarrow \mathscr{P}(E)$ of a data structure $D$ is a mapping such that  \\
         {\small
          $\sigma_D(x, root, e) = 
            \begin{cases}
            e' \mathit{if} \ Input(x) \land D(root,~e) \land \  D(root,~e') \land \ Constraints(root,~e,~e') \\
            \bot \ otherwise
            \end{cases}
          $
          } \\
         where $Constraints$ imposes the input-output relationship between $e$ and $e'$
         and $Input$ checks whether the input is well-formed. \\
        \textit{Definition} We say that $\sigma_D$ is computable in $n$ primitive steps if \newline 
         {\small
         \hphantom{th} $\forall x. \forall e. \ \sigma_D(x,~root,~e) = e' \neq \bot \Rightarrow   \newline 
         \hphantom{thequi}
         \exists x_1,\ldots,x_n,~
          y_1,\ldots,y_n,~\ell_1,..\ell_n.~s(x_n,y_n,\ell_n,s(x_{n-1},y_{n-1},\ell_{n-1},~s(..(s(x_1,~y_1,~\ell_1,~e)..))=e'$
          } \\ 
        We denote this relation as $\mathscr{C}_n = \{\sigma_{D} : \sigma_D \ is \ computable \ in \ n \ primitive \ steps \}$  \\ 
        We consider precisely the class $\mathscr{C}_n$ in this paper.
        For example, $insert_{list} \in \mathscr{C}_2$ and $delete_{ebst}\footnote{$list$ is synonymous with Linked Lists, $ebst$ with External BSTs} \in \mathscr{C}_1$.

     \subsection{Trace Correspondence and Program Extraction}
     To precisely capture the imperative program extraction from the answer sets, we define the following notions. 
     Let $M$ be an answer set of a normal logic program $\Pi$. Then, we define $T_M$ to be the sequence of all actions present in $M$. Let $L(T_M)$ denote length of the trace $T_M$.
      Literals of a model $M$ are all propositions that are true in $M$. A model $M$ can be treated as a set. \\ 
     \textit{Definition} Let $M_1,~M_2$ be answer sets of some program $\Pi$ and $T_{M_1},~T_{M_2}$, their corresponding traces. 
     We say that $M_1$ embeds ($\rightsquigarrow$) into $M_2$ if there is a structure preserving map of nodes in $M_1$ to nodes in $M_2$. Let $N_{M_1}$ and $N_{M_2}$ denote
     the nodes of models $M_1$ and $M_2$ respectively.  \\
     \textit{Definition} An embedding $f: N_{M_1}\hookrightarrow N_{M_2}$  is a mapping such that
     {\small
      \begin{align}
       \tag*{}
       \forall \ lit. : lit(x_1,~x_2, \ldots,~x_n) \in M_1 \quad \Rightarrow \quad lit(f(x_1),~f(x_2), \ldots,~f(x_n)) \ \in \  M_2 \\
       \tag*{}
      M_1 \; \rightsquigarrow \; M_2 \quad \Leftrightarrow \quad (\exists f.~f~:~N_{M_1} \; \hookrightarrow \; N_{M_2}) \ \land \ (\forall p. : p \ \in \  null_{M_1} \; \Rightarrow \; p \ \in \  M_2) 
         \end{align}
    }
    {\noindent}where $null_{M_1}$ is the set of nullary predicates of $\Pi$ that are true in $M_1$. \\
      \textit{Definition}
      We say that $M_1$ and $M_2$ have a trace correspondence ($\sim$) iff either $M_1$ embeds into $M_2$ or vice-versa, and both $T_{M_1}$ and $T_{M_2}$ have the same length. 
      {\small 
      \begin{align}
      \tag*{}
       M_1 \; \sim \; M_2 \quad \Leftrightarrow \quad ((M_1 \; \rightsquigarrow \; M_2) \ \lor \  (M_2 \; \rightsquigarrow \; M_1)) \ \land \  (L(T_{M_1}) \  = \ L(T_{M_2}))
       \end{align}
       }
      \textit{Few Notations} Let $\Pi_{\sigma}$ denote the ASP program encoding of some data structure D such that $\sigma \in \mathscr{C}_n$ for some constant $n$. Note that $\Pi_{\sigma}$ also encodes primitive step. $\Pi_{\sigma}[n+1]$ is the program $\Pi_{\sigma}$ with maximum allowed time $T = n+1$. It is clear that, $\Pi_{\sigma}[n+1]$ has models whose traces are of length $n$. 
     Let $\Pi^{sym}_{\sigma_{b}}[n+1]$ to represent the program that represents a run of $\Pi^{sym}_\sigma[n+1]$ using the base case definition of the data structure. Similarly, $\Pi^{sym}_{\sigma_{i}}[n+1]$ denotes the run of $\Pi^{sym}_\sigma$ using the inductive definition. Let $M_{b}$ and $M_{i}$ represent arbitrary models of $\Pi^{sym}_{\sigma_{b}}[n+1]$ and $\Pi^{sym}_{\sigma_{i}}[n+1]$ respectively. Let $M_\mathscr{B}, M_\mathscr{I}$ denote the set of all possible models of $\Pi^{sym}_{\sigma_{b}}[n+1]$ and $\Pi^{sym}_{\sigma_{i}}[n+1]$ respectively. 
     The following lemmas characterize the assumptions involved in the synthesis of the destructive update code.. \\
     \textit{Lemma 1.} The trace correspondence relation between two models is an equivalence relation.  (Trivial)
     \\     
     \textit{Lemma 2.}  If $M_{b} \sim M_{i}$, then there exists an imperative program satisfying $\Pi_{\sigma}$ \\
     \textit{Proof} Trace correspondence between $M_{b}$ and $M_{i}$ implies that the steps necessary to perform $\sigma$ are the equivalent under the isomorphism between the nodes involved in $M_{b}$ and $M_{i}$. By an inductive agreement, we can conclude that every data-structure instance constructed from the inductive definition has a trace correspondence with $M_{b}$. Consider an arbitrary model $M$ with no ground terms isomorphic to $M_{b}$. The nodes involved in $M$ can be represented by arbitrary names. Let $M$ be referred as the generic model of $\Pi_{\sigma}$. The generic model can be treated as the representative element of the equivalence relation. Now, the actions in $T_{M}$ ordered by their timestamp can be taken to be the primitive steps in imperative program $\Pi_{\sigma_{imp}}$ with the precondition consisting of conjunction of all literals that hold at time step $T = 0$ and facts that unconditionally hold true independent of time.  \\
     \textit{Note} It is assumed that there is a well-defined encoding of nodes, edges and keys in an imperative setting.

     \textit{Discussion} The imperative program constructed is the straight line program consisting of the actions satisfying the necessary \textit{goal}, sorted by time of occurrence. Complexity of computing the embedding between $M_b$ and $M_i$ is $\mathcal{O}(|M_\mathscr{B}|\times|M_\mathscr{I}|\times L^2)$ where $|M_\mathscr{B}|, |M_\mathscr{I}|$ represent cardinality of sets $M_\mathscr{B}, M_\mathscr{I}$ and $L^2$ is the number of mappings between traces of length $L$ belonging to models in $M_\mathscr{B}, M_\mathscr{I}$.  It is noted that $M_\mathscr{B}, M_\mathscr{I}$ may contain an exponential number of models in the size of all grounded literals in $\Pi_{\sigma}$. 
     \\
     \textit{Lemma 3.}   $\Pi^{sym}_{\sigma}$ preserves the models of $\Pi_{\sigma}$.  
      \\
       \textit{Proof} $\pisym$ has the following assumptions: Every ground symbol is unique. That is, no two ground symbols are equal. For numeric values, the inequality operator $<$ is the standard arithmetic inequality.
      $\pispec$ has facts corresponding to the ground symbols in $\pisym$ as follows:
      Every ground term is mapped to a unique symbol. If a symbol $x$ is a node, then the fact $eq\_node(x, x)$ is added for each $x$. For every other node symbol $y$, the fact $not\_eq\_node(x, y)$ is added as a fact. If symbols $x_1, x_2$ denote numbers, then $lt(x_1, x_2)$ is added as a fact if $x_1 < x_2$. Otherwise, $lt(x_2, x_1)$ is added. 
      Now we show that node equalities, inequalities and arithmetic inequalities from $\pispec$ are preserved in $\pisym$ \\ 
      Node equalities:  Only a ground term $x$ in $\pispec$ is equal to itself. Let $x'$ denote the mapping in $\pisym$. Because of the fact $eq\_node(x', x')$, the rule with head $not\_eq\_node(X, Y)$ is made defeasible for $x'$. Therefore, in no model $M'$ of $\pisym$, $not\_eq\_node(x', x')$ is inferred.
      A similar argument can be performed for node inequalities and arithmetic inequalities. 
      Let $\piinput$ denote the program consisting of just the input facts found in $\pispec$.
      Consider the fragment of rules denoted $\pifrag$ defining reflexivity, symmetry and transitive closure. 
      $\pisym$ can be stratified into two programs: $\pisym \setminus \pifrag$ and $\pifrag$. (Stratification is splitting of rules in a logic program into layers such that, rule heads defined in one layer are referred only in the body of rule heads in another layer). The rules in $\pifrag$ are not defined in $\pisym$. (They are only referred in $\pisym$) 
      By the splitting theorem \cite{lifschitz1994splitting}, the models of $\pisym \setminus \pifrag$ follow from the models of $\pifrag$. Consider all the rules in $\pisym \setminus \pifrag$. They are in one-one correspondence with $\pispec \setminus \piinput$. Therefore $\pisym \setminus \pifrag$ preserves the models of $\pispec \setminus \piinput$. Further, we have already proved that the node equalities, inequalities and arithmetic inequalities are preserved in $\pisym$. That is, they are also preserved in $\pifrag$. This implies the unique model of $\piinput$ is preserved by $\pifrag$. Combining the fragments, we have $(\pisym \setminus \pifrag) \cup (\pifrag) = \pisym$ preserves the models of $\pispec = (\pispec \setminus \piinput) \cup (\piinput)$ \\
     \textit{Lemma 4.} Partial Deduction is equivalent to complete deduction for Linked Lists and External BSTs. 
      \textit{Proof for Linked Lists} Consider the partially unfolded list in the inductive step: \texttt{edge(h, x). edge(x, y). key(h, kx). key(x, kx). key(y, ky). lt(kh, kx). lt(kx, ky). admissible(y).}
     Through arrow notation let the partial list be represented as $h \rightarrow x \rightarrow y$. Note that $y$ is suspended in $\pisym[t,k]$. From the partially unfolded list, it is clear that $y$ does not have any outgoing edge. Further due to rule $\mathit{Suspended-Unmodified}$, $y$ is not modified. That is, there is no \textit{link} operation affecting $y$. Without loss of generality, assume $Sat(\pisym[t], k)$ is true. 
Now, consider the fully unfolded list: $h \rightarrow x \rightarrow t$. Let the program encoding the fully unfolded list be denoted by $\pisymprime[t]$. We now prove that $Sat(\pisymprime[t])$ is true. Assume for the sake of contradiction, Sat($\pisymprime[t])$ is false. That is, $Unsat(\pisymprime[t])$ is true. Consider the node $t$ in $h \rightarrow x \rightarrow t$ (in $\pisymprime[t]$). $t$ is not a suspended node (as the list is fully unfolded in $\pisymprime[t]$). Consider, $\pisym[t,k]$ and $\pisymprime[t]$. $y$ and $t$ have the same key-ordering relative to other nodes in the list.  This is because both the lists $h \rightarrow x \rightarrow y$ and $h \rightarrow x \rightarrow t$ are isomorphic. $y$ is not modified in $\pisym[t, k]$. ($y$ is suspended). Although $t$ is not suspended, consider the constraint \textit{No-Node-Beyond-Tail}. $t$ cannot be modified in $\pisymprime[t]$. Therefore although $y$ and $t$ have different roles each in $\pisym[t, k]$ and $\pisymprime[t]$, they behave the same way. Rest of the list structure ($h \rightarrow x$) is the same in both $\pisym[t, k]$ and $\pisymprime[t]$. Therefore $Unsat(\pisym[t,k])$ is false. But we know that $Sat(\pisym[t, k])$ is true. Therefore Sat($\pisymprime[t]$) is true. The choice of $k$ is arbitrary and can be any integer. Therefore, Partial deduction is reduced to Complete Deduction. \\ 
     \section{Synthesis Procedure}
      We summarize the steps involved in the two procedures below. We refrain from referring the procedure to an algorithm as several conditions involved in the steps may not hold for an arbitrary pointer data structure.  $\Pi_{eq}$ denotes the equational theories capturing reflexive, symmetric, and transitive relations of $eq\_node$ and  $eq\_key$ predicates. Only transitivity is captured for the predicate $lt$. The function \textit{Models,} invokes an answer set solver to produce the models of an input program \cite{gelfond1988stable}. The function \textit{Get-Program-And-Precondition} sorts the primitive destructive pointer update steps and generates the necessary precondition.  The function \textit{Check-Traverse} checks whether the base case of the $traverse$ relation includes $Pre$. Otherwise, it performs a rewrite by appending $Pre$ as a conjunct in the base case of $traverse$. The programs $\pisym, \pitrav$ represent an arbitrary pointer-data structure and not just Linked Lists. 
        {
       \begin{algorithmic}[1]
            \setstretch{1.15}
            \Procedure{Synthesize-Pointer-Data-Structure}{$\pispec, \pitrav$} 
             \State ($M_{b}, M_{i})$ $\gets$ \textsc{Check-Sat-Upd($\pispec$)} \Comment{Inductive Proof for  Update}
             \If{\textsc{Has-Trace-Correspondence($M_{b}, M_{i}$)}}
                \State $M_{generic} \gets$ \textsc{Get-Generic-Model($M_b$)} \Comment{Only non-ground terms in trace}
                \State $(Update\_Func, Pre) \gets$ \textsc{Get-Program-And-Precondition($M_{generic}$)} 
                \Else 
                \  \textsc{ABORT}
            \EndIf
            \State  $Traverse \gets$  \textsc{Check-Sat-Trav($\pitrav, Pre$)} \Comment{Inductive Proof}
             \State  $Traverse\_Func \gets$ \textsc{Get-Recursive-Function($traverse$)}
             \State  $Imperative\_Code \gets Traverse\_Func \circ Update\_Func$ \Comment{Compose}
                 \State \Return $Imperative\_Code$
            \EndProcedure
            \\ 
            \Procedure{Check-Sat-Upd}{$\pispec$}
               \State $\pisym \gets$ \textsc{Abstract-Concrete-Operators($\pispec$)} 
               \State $\pisymprime \gets \pisymprime \cup \pieq$ 
               \State $\pisym \gets \pisymprimedeux \cup$ \textsc{Rewrite-Rules-Constraints($\pisymprimedeux$)} 
               \State $Sat \gets False, t \gets 0$
               \While{$Sat = False$} \Comment{Program search iterative in length of trace}
                   \If{$Sat(\pisym[t,0]) \land Sat(\pisym[t,k]) \land Sat(\pisym[t,k+1])$} 
                            \State    \Return ($True,$ \textsc{Models($\pisym[t,0]$),   Models($\pisym[t,k]$))}
                         \EndIf   \Comment{$Sat(\pisym[t,k]), Sat(\pisym[t,k+1)$: partial deduction}
                     \State $t \gets t + 1$        
               \EndWhile
            \EndProcedure
            \\ \\ \\
            \Procedure{Check-Sat-Trav}{$\pitrav, Pre$}
                  \State $\pisymtrav \gets$ \textsc{Abstract-Concrete-Operators($\pitrav$)}  \State $\pisymtravprime \gets \pisymtrav \cup \pieq$ 
                   \State $\pisymtravprimedeux \gets \pisymtravprime \cup$ \textsc{Rewrite-Rules-Constraints($\pisymtravprime$)}
                   \State $traverse \gets$ \textsc{Get-Traverse-Relation($\pisymtravprimedeux$)} 
                    \State $\pisymtrav \gets$ \textsc{Rewrite-Traverse-With-Pre($\pisymtravprimedeux,traverse,Pre)$}  
                     \If{$Sat(\pisymtrav, 0) \land Sat(\pisymtrav[k]) \land Sat(\pisymtrav[k+1])$} 
                            \State \Return $(True, traverse)$   \Comment{No time param for $\pisymtrav$}
                           \EndIf
                    \State \Return $(False, \bot)$
            \EndProcedure
       \end{algorithmic}
         }

      \section{Conclusion and Future Work}
       We have demonstrated that the insert operation for a linked list can be synthesized from an ASP specification. We can also similarly synthesize the code for delete operation. The goal simply changes to as shown below:
       {\small $\{goal(T) \; \leftarrow \; key(X, K), target(K), not \ reachable(X, T)\}$},
       while relaxing the constraint \textit{No-Key-Loss} to affect only non-target keys. To demonstrate the generality of our technique, the same approach applied to External-BSTs is included in Appendix A. In defining the goal for delete operation, all that was needed was a simple change from $reachable(X, T) \ to \ not \ reachable(X, T)$. Further, the rules in the world of linked lists are written naturally, in almost a common-sensical fashion. Setting aside the superficial burden of not having to write quantifiers, the rules represent succinct definitions of a subject theory. Negative concepts such as \textit{unreachability} are naturally qualified with \textit{NAF}. Without using NAF, providing a definition of $unreachable$ is not straightforward, and has to be defined procedurally. Therefore, not relying on hard evidence for negative information makes ASP a better formalism than First-Order Logic, to write specifications. Note that this comes at a price of causality. That is, there has to be a strong causal relation between a rule definition and its consequent. For example, rule $\{p \; \leftarrow \; q \}$ should mean $\{p \leftrightarrow q \}$. Otherwise, soundness of the judgement through negation-as-failure would be compromised. 
       \par Our work relates with several ideas existing in Program Analyses, Logical Formalisms and Transformation. Program Analyses of Heap manipulating programs have been studied well, which introduce new formalisms to capture in hindsight, the semantics involved in heap-modifying program steps. Most popular among them is Separation logic an extension of First-Order Logic. Three-valued logics \cite{sagiv2002parametric} have also been used to analyse heap-manipulating programs. Other significant shape analyses use predicate abstraction \cite{balaban2005shape}. We believe that a semantics-guided approach to program transformation should be sufficient for the program synthesis problem. In spirit, our work is closely related to the semantics preserving transformation due to Darlington \cite{darlington1981experimental} and extraction of program and proofs from axiomatic descriptions of data structures due to Manna \cite{manna1971toward}. When compared to Abstract Interpretation, our procedure retains precision when performing partial evaluation (deduction). The idea that Partial Deduction performs an inductive proof is also detailed elsewhere \cite{lehmann2003inductive}. We have applied it independently in ASP to pointer data structures. Our procedure currently generates straight line programs for destructive update and if-else branches only for recursive definitions. Further, we require that the number of primitive operations involved in the destructive update for an algebraic operation must be constant (the class $C_n$). Because both the destructive update and traversal part are verified for all instances of the data structure (inductively), termination of the generated program is guaranteed for well-formed inputs. Much work still remains, to generalize to more classes of pointer-data structures such as Internal BSTs and more algebraic operations such as Linked List reversal. 
       %GG:
       Eventually, our goal is to automatically generate programs for manipulating  \textit{concurrent} data structures.
       
      \printbibliography
      
\section*{Appendix A: Synthesis for External BSTs}
 An External BST is a binary search tree with the keys of the tree present only at the leaves. All the internal nodes are used for ``routing" purposes. Every internal node has two children: left-child and right-child
 Nodes on the heap are modelled by the \textit{node} relation. Edges in the tree are distinguished by the \textit{left} and \textit{right} relations. The relation \textit{left(X, Y)} denotes node \textit{Y} is the left child of node \textit{Y}. The relation \textit{right(X, Y)} is defined similarly. Every leaf node has two ``nil" nodes as its children. 
 
 \subsection*{Recursive Tree Definition}
  The recursive tree definition is given by the \textit{tree} relation as follows. There is a designated node called ``root"" represented by the relation \textit{root}.
  \begin{align}
      \tag{Tree-Root}
      tree(T) \; \leftarrow \; root(X, T), \textit{left(X, Y, T)}, right(X, Z, T), tree(X, T), tree(Y, T) \\
      \tag{Tree-Recursive}
      tree(X, T) \; \leftarrow \; X \neq nil, Y \neq nil, Z \neq nil, \textit{left(X, Y, T)}, right(X, Z, T), tree(Y, T), tree(Z, T) \\
      \tag{Tree-Leaf}
      tree(X, T) \; \leftarrow \; X \neq nil,  \textit{left(X, nil, T)}, right(X, nil, T) \\
      \tag{Tree-Constraint}
      \; \leftarrow \; not \ tree(T)
  \end{align}
  The variable $T$ is understood to be time step. 
  
  \subsection*{Admissibility}
  The External BST also requires a key-ordering of nodes. Nodes with keys less than the root are part of the left sub-tree and with keys greater than that of root are part of the right sub-tree. We give this definition as follows: There cannot exist a node in the left sub-tree with a key-value greater than the root. If such a node exists, then it represents an unsafe tree. Our definition of admissible uses \textit{negation-as-failure} over unsafe trees. To define unsafe trees, we need notions of \textit{reachable} and \textit{descendant}. They are used in the usual sense. That is, \textit{reachable(X)} denotes that the node \textit{X} is reachable from the root. Whereas \textit{descendant(X, Y)} denotes that node \textit{Y} is the descendant of node \textit{Y}. We also distinguish between left and right descendants using \textit{descendant(X, Y, left)} and \textit{descendant(X, Y, right)} respectively. 
  To define descendant, we need the abstraction of \textit{child(X, Y)} which denotes node \textit{Y} is either a left-child or a right-child of \textit{X}. 
  
  \begin{alignat}{1}
      \tag{Child-Left}
       child(X, Y, T) \; \leftarrow \; \textit{left(X, Y, T)} & \\
      \tag{Child-Right}
      child(X, Y, T) \; \leftarrow \; right(X, Y, T) & \\ 
      \tag{Descendant-Left-Base}
      descendant(X, Y, \textit{left}, T) \; \leftarrow \; \textit{left(X, Y, T)} & \\
      \tag{Descendant-Right-Base}
      descendant(X, Y, right, T) \; \leftarrow \; right(X, Y, T) & \\
      \tag{Descendant-Left-Recursive}
      descendant(X, Y, \textit{left}, T) \; \leftarrow \; \textit{left(X, Z, T)}, descendant(Z, Y, T) \\
      \tag{Descendant-Right-Recursive}
      descendant(X, Y, right, T) \; \leftarrow \; right(X, Z, T), descendant(Z, Y, T)
  \end{alignat}
  
  \subsubsection*{Reachability}
     \begin{align}
         \tag{Reachable-Root}
         reachable(X, T) \; \leftarrow \; root(X, T) \\
         \tag{Reachable-Recursive}
         reachable(X, T) \; \leftarrow \; root(Y, T), descendant(Y, X, T)
     \end{align}
  
  \subsubsection*{Unsafe Trees}

  \begin{alignat}{1}
       \tag{Unsafe-Subtree-Left}
       \textit{unsafe(T)} \; \leftarrow \; reachable(X, T), descendant(X, Y, T, \textit{left}), key(X, K_1), key(Y, K_2), K_1 < K_2 \\
       \tag{Unsafe-Subtree-Right}
       \textit{unsafe(T)} \; \leftarrow \; reachable(X, T), descendant(X, Y, T, right), key(X, K_1), key(Y, K_2), K_1 > K_2 \\
       \tag{Unsafe-Leaf-Left}
       \textit{unsafe(T)} \; \leftarrow \; leaf(X, T), not \ left(X, nil, T) \\
       \tag{Unsafe-Leaf-Right}
       \textit{unsafe(T)} \; \leftarrow \; leaf(X, T) not \ right(X, nil, T)
  \end{alignat}
 
The relation \textit{Leaf(X, T)} denotes that node \textit{X} is a leaf node and its definition is given below:
  \begin{align}
       \tag{Leaf}
       \textit{leaf(X, T)} \; \leftarrow \; reachable(X, T), \textit{left(X, nil, T)}, right(X, nil, T)
  \end{align} 
Finally, \textit{admissible} is simply \textit{unsafe} qualified with negation-as-failure.
  \begin{align}
     \tag{Admissible}
      admissible(T) \; \leftarrow \; not \ unsafe(T) \\
      \tag{Admissible-Constraint}
      \; \leftarrow \; not \ admissible(T)
  \end{align}
  
 Two primitive operation supported are \textit{link-left} and \textit{link-right} which correspondingly link the left and right pointers of a node to another node. For example, \textit{link-left(x, y)} links the left pointer of $x$ to $y$. In modern imperative languages, this is similar to the statement: $x.left = y$.  
 The actions are guessed through circular negation as before. The effects of the actions are also given below:
 \begin{align}
     \tag{Link-Left-Guess}
     \textit{link-left(X, Y, T)} \; \leftarrow \; not \ \textit{neg-link-left(X, Y, T)} \\
     \tag{Neg-Link-Left}
     \textit{neg-link-left(X, Y, T)} \; \leftarrow \; not \ \textit{link-left(X, Y, T)} \\
     \tag{Link-Right-Guess}
     \textit{link-right(X, Y, T)} \; \leftarrow \; not \ \textit{neg-link-right(X, Y, T)}  \\
     \tag{Neg-Link-Right}
     \textit{neg-link-right(X, Y, T)} \; \leftarrow \; not \ \textit{link-right(X, Y, T)}  \\
     \tag{New-Left-Child}
     \textit{left(X, Y, T+1)} \; \leftarrow \; \textit{link-left(X, Y, T)} \\
     \tag{New-Right-Child} 
     \textit{right(X, Y, T+1)} \; \leftarrow \; \textit{link-right(X, Y, T)} 
 \end{align}
 
A node should retain the same left (right) child when not modified. We first define the notion of \textit{modified} and use NAF over \textit{modified} to represent retention of previous left (right) child.

\begin{align}
    \tag{Modified-by-Link-Left}
    modified(X, T) \; \leftarrow \; \textit{link-left}(X, Y, T) \\
    \tag{Modified-by-Link-Right}
    modified(X, T) \; \leftarrow \; \textit{link-right}(X, Y, T) \\
    \tag{Same-Left-Child}
    \textit{left}(X, Y, T+1) \; \leftarrow \; \textit{left}(X, Y, T), not \ modified(X, T) \\
    \tag{Same-Right-Child}
    right(X, Y, T+1) \; \leftarrow \; right(X, Y, T), not \ modified(X, T)
\end{align}
 
 \subsubsection*{Constraints on Executability of Pointer Linkage}
  \ The actions \textit{link-left, link-right} cannot execute at the same time. We restrict the using constraints limiting their simultaneuous execution. Similar to Linked List \textit{link} operation, only a single node can be modified at any point in time. 
  
  \begin{alignat}{1}
      \tag{Link-Actions-Mutex}
       & \; \leftarrow \; \textit{link-left(X, Y, T)}, \textit{link-right}(X', Y', T) \\
       \tag{Link-Left-Mutex}
       & \; \leftarrow \; \textit{link-left(X, Y, T)}, \textit{link-left}(X, Y', T), Y  \neq Y' \\
       \tag{Link-Right-Mutex}
       & \; \leftarrow \;  \textit{link-right}(X, Y, T), \textit{link-right}(X, Y', T), Y \neq Y' \\
       \tag{Link-Single-Node-1}
       & \; \leftarrow \; \textit{link-left(X, Y, T)}, \textit{link-left}(X', Z, T), X \neq X' \\
       \tag{Link-Single-Node-2}
       & \; \leftarrow \; \textit{link-right}(X, Y, T), \textit{link-right}(X', Z, T), X \neq X'
  \end{alignat}
  
  \subsubsection*{Constraints on Heap-Structure}
     \ A node may not have same node as both left and right child. Two nodes should not be descendants of each other. 
        \begin{align}
            \tag{No-Duplicate-Child}
            \; \leftarrow \; \textit{left}(X, Y, T), right(X, Y, T) \\
            \tag{No-Mutual-Descendants}
            \; \leftarrow \; descendant(X, Y, T), descendant(Y, X, T) \\
            \tag{No-Left-Selfloop}
            \; \leftarrow \; \textit{left}(X, X, T) \\
            \tag{No-Right-Selfloop}
            \; \leftarrow \; \textit{right(X, X, T)}
        \end{align}
   \subsubsection*{Presence of Keys in External BST}
       \  A key is present in an External BST iff there is a reachable leaf node containing the same key.
        Additionally for External BSTs, every internal node should have two children.
        We make the notion of internal and external nodes explicit below:
        \begin{alignat}{1}
            \tag{External-Node}
            & external\_node(X, T) \; \leftarrow \; reachable(X, T), \textit{leaf}(X, T) \\
            \tag{Internal-Node}
            & internal\_node(X, T) \; \leftarrow \; reachable(X, T), not \ external\_node(X, T) \\
            \tag{Key-Present}
            & present(K, T) \; \leftarrow \; external\_node(X, T), key(X, K) \\ 
            \tag{Has-Left-Child}
            & \textit{has\_left\_child(X, T)} \; \leftarrow \; X \neq nil, left(X, Y, T) \\
            \tag{Has-Right-Child}
            & \textit{has\_right\_child(X, T)} \; \leftarrow \; X \neq nil, right(X, Y, T)  \\
            \tag{Internal-Node-Left-Child}
            & \; \leftarrow \; internal\_node(X, T) \; \leftarrow \; not \ \textit{has\_left\_child}(X, T) \\
            \tag{Internal-Node-Right-Child}
            & \; \leftarrow \; internal\_node(X, T) \; \leftarrow \; not \ \textit{has\_right\_child}(X, T)
        \end{alignat}
        
        Internal nodes continue to be internal nodes their entire lifetime (unless removed from tree). Similarly every initial reachable node should continue to be reachable. 
        \begin{align}
            \tag{Internal-Node-Invariant}
            \; \leftarrow \; internal\_node(X, 0), not \ internal\_node(X, T) \\
            \tag{Reachable-Invariant}
            \; \leftarrow \; reachable(X, 0), not \ reachable(X, T)
        \end{align}
        
        No keys should be lost as part of the insert operation.
        \begin{align}
            \tag{No-Key-Loss}
            \; \leftarrow \; present(K, 0), not \ present(K, T)
        \end{align}
        
        Finally, we state the objective: target key must be present in the tree
        
        \begin{align}
            \tag{Objective}
            goal(T) \; \leftarrow \; target\_key(K), not \ present(K, 0), present(K, T)
        \end{align}
        
        \subsection*{Partial Deduction in External BST}
        
        We use the inductive definition of \textit{tree} predicate to perform partial evaluation. The induction performed is on the height of the tree. The trees for base case and inductive cases are shown below:
        
        \newpage
        \begin{mdframed}[leftmargin=10pt,rightmargin=10pt]
          Base case: $tree(0) \; \leftarrow \; root(x, 0), left(x, y, 0), right(x, z, 0), left(y, nil, 0), \newline \hphantom{thequickbrownfoxju} right(y, nil, 0), left(z, nil, 0), right(z, nil, 0)$ \\ 
          Inductive cases: \\
          $tree(0) \; \leftarrow \; x \neq nil, y \neq nil, z \neq nil, root(x,0), \textit{left}(x, y, 0), right(x, z, 0), \newline \hphantom{thequickbr} \textit{left}(y, nil, 0),  right(y, nil, 0), \textit{\textbf{tree(z, 0)}}$ \\
          $tree(0) \; \leftarrow \; x \neq nil, y \neq nil, z \neq nil, root(x, 0),  \textit{left}(x, y, 0), right(x, z, 0), \newline \hphantom{thequickbr} \textit{\textbf{tree(y, 0)}}, \textit{left}(z, nil, 0), right(z, nil, 0)$ \\
          $tree(0) \; \leftarrow \; x \neq nil, y \neq nil, z \neq nil, root(x, 0), \textit{left}(x, y, 0), right(x, z, 0),
          \newline \hphantom{thequickbr} \textit{left}(y, nil, 0), right(y, nil, 0), \textit{left}(z, y', 0), right(z, z', 0), y' \neq nil,
          \newline \hphantom{thequickbr} z' \neq nil, \textit{left}(y', nil, 0), right(y', nil, 0), \textit{\textbf{tree(z', 0)}}$
        \end{mdframed}
        
        The predicates shown in bold are suspended and the nodes involved in the predicates are suspended. 
        For example in the first inductive case, the node \textit{z} is suspended. Similarly, \textit{y, z'} are suspended in the remaining two cases. Key-ordering is not present in the inductive tree definition unlike the inductive definition of Linked lists. The key constraints are automatically enforced through the \textit{unsafe} predicate. They are not easily seen but are captured in the theory.
    
        \subsection*{Synthesizing Destructive Update for External BST Insert operation}
             The insert operation of an External BST happens at the leaf nodes. As part of the insert operation, two new nodes are inserted into the BST. One node is the target node where as an additional new node is created to properly perform linkage. We assume this additional new internal node is provided by the domain expert. Let this node be denoted by the fact \textit{node(internal)}. Let the target node be represented by the fact \textit{node(target)}
             The set of input facts for one variation of base case are shown below:
             \begin{mdframed}[leftmargin=10pt,rightmargin=10pt]
              $root(a, 0). \ left(a, b, 0). \ right(a, c, 0). \ left(b, nil, 0). \ right(b, nil, 0), \newline left(b, nil, 0). \  right(c, nil, 0). \ key(a, k_a). \ key(b, k_b). \ key(c, k_c). \ key(target, k_t). \newline \ key(internal, k_i). \ lt(b, a). \ lt(a, c). \ lt(k_c, k_t). 
              $
             \end{mdframed}
             The corresponding linkage of pointers is shown below:
             \begin{mdframed}[leftmargin=10pt, rightmargin=10pt]
             $\textit{link-right}(internal,c,0). \ \ \textit{link-left}(internal,target,1). \ \ \textit{link-right(a,internal,2)}$
             \end{mdframed}
             Next page visualizes the target node insertion into the EBST. Round nodes are internal nodes while square nodes are external. The nodes are annotated by the key value they carry. There are three more variations of the base case,
             \begin{enumerate}
                 \item $lt(k_a, k_t), lt(k_t, k_c)$
                 \item $lt(k_t, k_a), lt(k_t, k_b)$
                 \item $lt(k_t, k_a), lt(k_b, k_t)$
             \end{enumerate}
             The other three cases can be visualized similar to the figure shown.

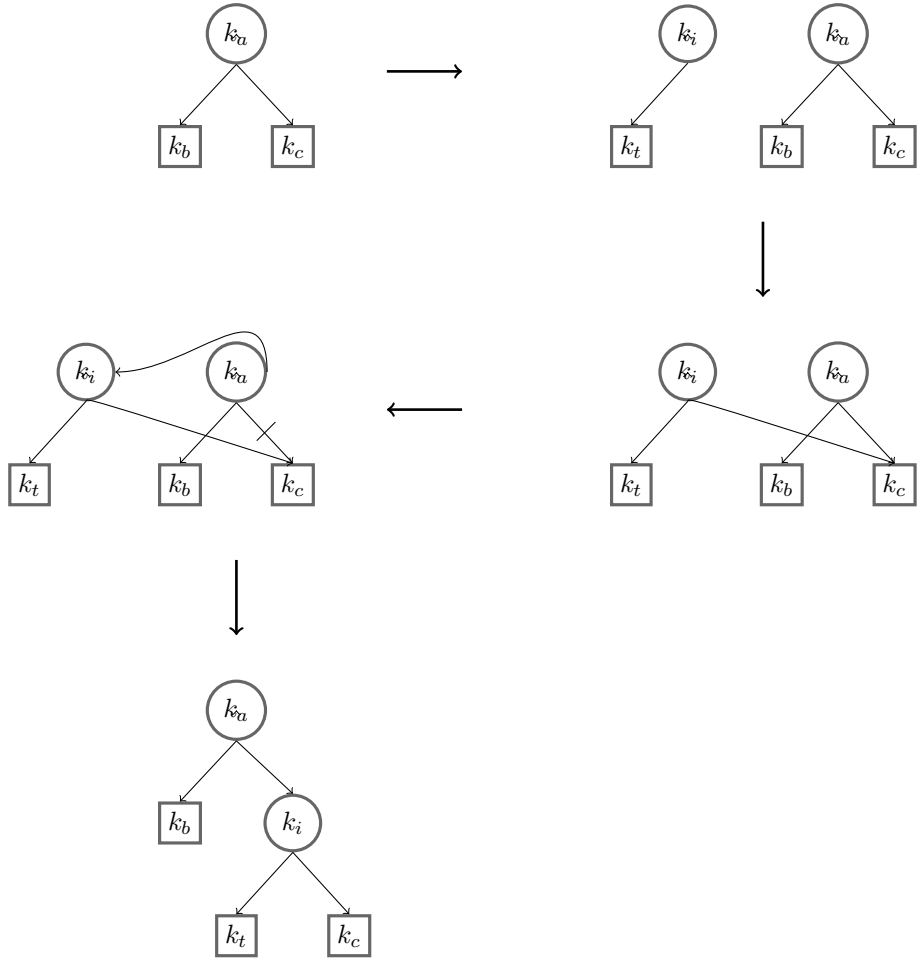
\begin{center}
\begin{tikzpicture}[
roundnode/.style={circle, draw=black!60,  very thick, minimum size=7mm},
squarednode/.style={rectangle, draw=black!60,  very thick, minimum size=5mm},
->
]

%Nodes
%\node[squarednode]      (maintopic)                              {2};
%\node[roundnode]        (uppercircle)       [above=of maintopic] {1};
%\node[squarednode]      (rightsquare)       [right=of maintopic] {3};
%\node[roundnode]        (lowercircle)       [below=of maintopic] {4};
 
 \draw (-10,0) node[roundnode]  {$k_a$}
    child {node[squarednode] {$k_b$}}
    child {node[squarednode]  {$k_c$}};
\draw[line width=1pt, rounded corners=1pt] (-8,-0.5) -- (-7, -0.5);
%\draw[line width=1pt, rounded corners=1pt] (-7, -0.5) -- (-7.25,0.25-0.5);
%\draw[line width=1pt, rounded corners=1pt] (-7, -0.5) -- (-7.25,-0.25-0.5);
 \draw  (-4,0) node[roundnode]  {$k_i$}
     child {node[squarednode]  {$k_t$}}
     child [missing];
 \draw (-2,0) node[roundnode]  {$k_a$}
    child {node[squarednode] {$k_b$}}
    child {node[squarednode]  {$k_c$}};
      
 \draw[line width=1pt, rounded corners=1pt] (-3,-0.5-2) -- (-3, -0.5-3);

 \draw  (-4,-4.5) node[roundnode] (i) {$k_i$}
     child {node[squarednode]  {$k_t$}}
     child [missing];
 \draw (-2,-4.5) node[roundnode]  {$k_a$}
    child {node[squarednode] {$k_b$}}
    child {node[squarednode] (c) {$k_c$}};
 \draw[->] (i.south) .. controls +(up:0.5mm) .. (c.north);
 
 \draw  (-12,-4.5) node[roundnode] (i1) {$k_i$}
     child {node[squarednode]  {$k_t$}}
     child [missing];
 
 \draw (-10,-4.5) node[roundnode] (a1)  {$k_a$}
    child {node[squarednode] {$k_b$}}
    child {node[squarednode] (c1) {$k_c$}
    };
  \path (a1) -- (c1) node[draw,strike out, pos=0.5] {};
 
  \draw[->] (i1.south) .. controls +(up:0.5mm) .. (c1.north);
  \draw[->] (a1.east) .. controls +(up:12mm) and +(right:10mm) .. (i1.east);

 \draw[line width=1pt, rounded corners=1pt]  (-7, -5) -- (-8,-5) ;

  \draw[line width=1pt, rounded corners=1pt] (-10,-7) -- (-10, -8);
  
   \draw (-10,-9) node[roundnode]   {$k_a$}
    child {node[squarednode] {$k_b$}}
    child {node[roundnode]  {$k_i$}
          child {node[squarednode] {$k_t$}}
          child {node[squarednode] {$k_c$}}
    };
                  
\end{tikzpicture}
\end{center}             
              The four answer sets for base case are shown below. Node \textit{internal} is written as node \textit{i} for short. Similarly, node \textit{target} is written as node \textit{t} for short.
    \begin{mdframed}[leftmargin=10pt,rightmargin=10pt]
    {\small
     $Case \ 1:  lt(k_a, k_t). \ lt(k_t, k_c). \ \textit{link-left}(i, t, 0). \  \textit{link-right}(i, c, 1). \ \textit{link-right}(a, i, 2). \newline  
     Case \ 2:  lt(k_a, k_t). \ lt(k_c, k_t). \ \textit{link-right}(i, t, 0). \
     \textit{link-left}(i, c, 1) \ \textit{link-right}(a, i, 2). \newline
     Case \ 3: lt(k_t, k_a). \ lt(k_t, k_b). \ \textit{link-left}(i, t, 0) \ 
     \textit{link-right}(i, b, 1). \ \textit{link-left}(a, i, 2). \newline
     Case \ 4: lt(k_t, k_a). \ lt(k_b, k_t). \ \textit{link-right}(i, t, 0). \ 
     \textit{link-left}(i, b, 1). \ \textit{link-left}(a, i, 2).  
     $
     }
    \
    \end{mdframed}
     Note that there are 4 generic models for External BSTs as opposed to just one generic model for Linked List which implies there 4 different preconditions for traversal to compose with the destructive update.
     
     \subsection*{Traversal Relation for External BST}
      There are 4 different preconditions synthesized from the destructive update due to the 4 variations a target node might be inserted into the tree.
      The precondition for the entire insert operation is therefore a disjunction of the preconditions of the 4 generic models.
    $Pre(X, Y) \; \equiv \; Pre1(X, Y) \lor Pre2(X, Y) \lor Pre3(X, Y) \lor Pre4(X, Y)$  \\ \\
    $Pre1(X, Y) \; \equiv \; reachable(X) \ \land \ reachable(Y) \ \land \ descendant(X, Y) \ \land \ \textit{left}(X, Y)    
    \newline  \hphantom{thequickbro} \land
          \ internal\_node(X) \ \land \ external\_node(Y) \ \land \ key(X, K_X) \ \land key(Y, K_Y) \newline
           \hphantom{thequickbro} \land  \ external\_node(Y) \ \land \ key(X, K_X) \ \land key(Y, K_Y) \ \land \ key(target, K_{target})
           \newline \hphantom{thequickbro} \land \ lt(K_Y, K_X) \ \land \ lt(k_{target}, K_Y) \ \land \ reachable(Z) \ \land \ descendant(X, Z) 
           \newline \hphantom{thequickbro}  \land \ right(X, Z) \ \land \ key(Z, K_Z) \ \land \ lt(K_X, K_Z)$ \\ \\
    $Pre2(X, Y) \; \equiv \; (Pre1(X, Y) \setminus \{lt(K_{target}, K_Y)\} \cup \{lt(K_Y, K_{target})\}$ \\ \\
    $Pre3(X, Y) \; \equiv \; reachable(X) \ \land \ reachable(Y) \ \land \ descendant(X, Y) \ \land \ \textit{right}(X, Y)  
    \newline  \hphantom{thequickbro} \land \
           internal\_node(X) \ \land \ external\_node(Y) \ \land \ key(X, K_X) \ \land \ key(Y, K_Y) \newline
           \hphantom{thequickbro} \land \ external\_node(Y) \ \land \ key(X, K_X) \ \land \ key(Y, K_Y) \ \land \ key(target, K_{target})
           \newline \hphantom{thequickbro} \land \ lt(K_X, K_Y) \ \land \ lt(k_{target}, K_Y) \ \land \ reachable(Z) \ \land \ descendant(X, Z) 
           \newline \hphantom{thequickbro} \land  \ \textit{left}(X, Z) \ \land \ key(Z, K_Z) \ \land \ lt(K_Z, K_X)$ \\ \\
    $Pre4(X, Y) \; \equiv \; (Pre3(X, Y) \setminus \{lt(K_{target}, K_Y)\} \cup \{lt(K_Y, K_{target})\}$  \\ \\
     The traversal relation is defined as follows:
     \begin{align}
         \tag*{}
         start\_traversal \; \leftarrow \; root(X), K_X > K_{target}, \textit{left}(X, Y), traverse(X, Y) \\ 
         \tag*{}
         start\_traversal \; \leftarrow \; root(X), K_X < K_{target},
         right(X, Y), traverse(X, Y) \\
         \tag*{}
         traverse(X, Y) \; \leftarrow \; Pre(X, Y) \\
         \tag*{}
         traverse(X, Y) \; \leftarrow \; child(X, Y), K_Y > K_{target}, \textit{left}(Y, Z), traverse(Y, Z) \\
         \tag*{}
         traverse(X, Y) \; \leftarrow \; child(X, Y), K_Y < K_{target}, right(Y, Z), traverse(Y, Z)
     \end{align} \\
     
     The synthesized imperative code would look as follows: s
    \begin{multicols}{2}
       \fontsize{8}{9}\selectfont
       \begin{algorithmic}[0]
       \Function{insert}{$target$}
       \State $(x,~y) \gets$ traverse($target$)
       \If{x.left = y \textbf{and} x.key $>$ target.key \textbf{and} target.key $<$ y.key}
       \State $internal.left \gets target$
       \State $internal.right \gets y$
       \State $x.left \gets internal$
       \EndIf
       \If{x.left = y \textbf{and} x.key $>$ target.key \textbf{and} target.key $>$ y.key}
       \State $internal.right \gets target$
       \State $internal.left \gets y$
       \State $x.left \gets internal$
       \EndIf
       \If{x.right = y \textbf{and} x.key $<$ target.key \textbf{and} target.key $<$ y.key}
       \State $internal.left \gets target$
       \State $internal.right \gets y$
       \State $x.right \gets internal$
       \EndIf
       \If{x.right = y \textbf{and} x.key $<$ target.key \textbf{and} target.key $>$ y.key}
       \State $internal.right \gets target$
       \State $internal.left \gets y$
       \State $x.right \gets internal$
       \EndIf
       
       \EndFunction
       \end{algorithmic}
       
       \begin{algorithmic}[0]
       \Function{traverse}{$target$}
       \If{target.key $<$ root.key}
       \State \Return traverse(root, root.left, target)
       \EndIf
       \State \Return traverse(root, root.right, target)
       \EndFunction
       \end{algorithmic}
       
       \begin{algorithmic}[0]
       \Function{traverse}{$x$, $y$, $target$}
          \If{Pre(x, y)}  /*Pre must be in residual form without reachable, descendant*/
          \State \Return $(x,~y)$
       \EndIf
       \If{target.key $<$ y.key}
       \State \Return traverse($y$, $y.left$, $target$)
       \EndIf
       \State \Return traverse($y$, $y.right$, $target$)
       \EndFunction
       \end{algorithmic}
       \end{multicols}

 \subsection*{Rewrites of Rules, Constraints for External BSTs}    
   \begin{alignat}{1}
        \tag{Rewrite-No-Left-Self-Loop}
    &    \; \leftarrow \; \textit{\textbf{not suspended(X), \ eq\_node(X, Y)}}, \textit{left(X, Y, T)} \\
       \tag{Rewrite-No-Right-Self-Loop}
    &   \; \leftarrow \; \textit{\textbf{not suspended(X), eq\_node(X, Y)}},  right(X, Y, T) \\
        \tag{Rewrite-Unsafe-Subtree-Left}
    &    \textit{unsafe(T)} \; \leftarrow \; reachable(X,~T), descendant(X,~Y,\textit{left},T),key(X,K_1),key(Y,K_2), \textit{\textbf{$lt(K_1,K_2)$}} \\
        \tag{Rewrite-Unsafe-Subtree-Right}
     &   \textit{unsafe(T)} \; \leftarrow \; reachable(X,~T), descendant(X,~Y,right,T),key(X,K_1),key(Y,K_2),\textit{\textbf{not\_lt($K_1,K_2)$}}  \\
        \tag{Reachable-Eq}
     &   reachable(X, T) \; \leftarrow \;  eq\_node(X, Y), reachable(Y, T) \\
        \tag{Descendant-Eq-1}
     &   descendant(X, Y, T) \; \leftarrow \; eq\_node(X, Z), descendant(Z, Y, T) \\
        \tag{Descendant-Eq-2}
     &   descendant(X, Y, T) \; \leftarrow \; eq\_node(Y, Z), descendant(X, Z, T) \\
        \tag{Left-Eq-1}
     &   \textit{left}(X, Y, T) \; \leftarrow \; eq\_node(X, Z), \textit{left}(Z, Y, T) \\
        \tag{Left-Eq-2}
     &   \textit{left}(X, Y, T) \; \leftarrow \; eq\_node(Y, Z), \textit{left}(X, Z, T) \\
        \tag{Right-Eq-1}
     &   right(X, Y, T) \; \leftarrow \; eq\_node(X, Z), right(Z, Y, T) \\
        \tag{Right-Eq-2}
     &   right(X, Y, T) \; \leftarrow \; eq\_node(Y, Z), right(X, Z, T) \\
        \tag{Has-Left-Child-Eq}
     &   has\_\textit{left}\_child(X, T) \; \leftarrow \; eq\_node(X, Y) has\_\textit{left}\_child(Y, T) \\
        \tag{Has-Right-Child-Eq}
     &   has\_right\_child(X, T) \; \leftarrow \; eq\_node(X, Y), has\_right\_child(Y, T) \\
        \tag{Leaf-Eq}
     &   \textit{leaf}(X, T) \; \leftarrow \; eq\_node(X, Y), \textit{leaf}(Y, T) \\ 
        \tag{External-Node-Eq} 
     &  external\_node(X, T) \; \leftarrow \; eq\_node(X, Y), \textit{leaf}(Y, T) \\
        \tag{Internal-Node-Eq}
     &   internal\_node(X, T) \; \leftarrow \; eq\_node(X, Y), internal\_node(Y, T)
    \end{alignat}

      \textit{Lemma 4} Partial deduction is equivalent to complete deduction 
      
      \textit{Proof for External BSTs} Let $\pisymbst[t]$ represent the encoding of External BST in this context. Here, the induction is on the depth $k$ of the tree. Consider the partially unfolded trees.
     
     \begin{mdframed}[leftmargin=10pt,rightmargin=10pt]
          Inductive cases: \\
          $tree \; \leftarrow \; x \neq nil, y \neq nil, z \neq nil, root(x) \  \textit{left}(x, y), right(x, z), \textit{left}(y, nil), \newline \hphantom{thequickbr}right(y, nil), \textit{\textbf{tree(z)}}$ \\
          $tree \; \leftarrow \; x \neq nil, y \neq nil, z \neq nil, root(x) \  \textit{left}(x, y), right(x, z), \textit{\textbf{tree(y})} \newline \hphantom{thequickbr}\textit{left}(z, nil), right(z, nil)$ \\
        \end{mdframed}

     Consider the first inductive definition. Let $\pisymbst[t]$ represent the encoding of partially unfolded tree. Node $z$ is suspended in $\pisymbst[t]$. Without loss of generality, assume $Sat(\pisymbst[t, k])$ is satisfiable. Here, $k$ denotes that the External BST has depth $k$. Now, construct a new tree as follows: \\
     $\{root(x) \ \textit{left}(x, y) \ right(x, z') \ \textit{left}(y, nil) \ right(y, nil) \ \textit{left}(z', nil) \ right(z', nil)\}$
     Let $\pisymprimebst[t]$ encode the newly constructed tree. It is clear that the key-ordering is isomorphic in both $\pisymbst[t, k]$ and $\pisymprimebst[t]$. 
     Now we prove that $Sat(\pisymprimebst[t])$ is true. Assume for the sake of contradiction, $Sat(\pisymprimebst[t])$ is false. That is, $Unsat(\pisymprimebst[t])$ is true. From $Sat(\pisymbst[t, k])$ and node $z$ suspended in $\pisymbst[t, k]$, we have $k_{target} < k_x$. That is, the target node is inserted into the left sub-tree of $x$. As the newly constructed tree is isomorphic to the old tree, $k_{target} < k_x$ is true in $\pisymprimebst[t]$. Therefore, node $z'$ is never modified in $\pisymprimebst[t]$. Thus, nodes $z$ and $z'$ behave the same way in $\pisymbst[t, k]$ and $\pisymprimebst[t]$ respectively. This implies, the left sub-tree is isomorphic in both $\pisymbst[t, k]$ and $\pisymprimebst[t]$. Therefore, $Unsat(\pisymbst[t, k])$ is true, which is a contradiction. Therefore, $Sat(\pisymprimebst[t])$ is true.

     \section*{Appendix B: Proof of Lemma 3 for LOPSTR 2020 Paper}
           \noindent {\bf \textit{Lemma 3.}} Partial deduction is equivalent to complete deduction for tree-based inductive data structures with suspended nodes. \\
        \noindent \textbf{\textit{Proof}}  Let $\Delta$ be an inductive data structure and $\delta'_{b+1}$ be the partial instance such that $\Pi_{\delta_b}$ is satisfiable and $\Pi_{\delta'_{b+1}}$ is satisfiable. The requirements of algebraic operation $\sigma$ are assumed to be encoded in $\Pi_{\delta_b}$ and $\Pi_{\delta'_{b+1}}$. Now for every complete instance $\delta_{b+i}$, $i = 1, 2, 3, ..$  we can make the following construction:
        Consider some instance $\delta_{b+i}$. Note that $\delta_{b} \in S_{\Delta}$ and $\delta_{b_i} \in S_{\Delta}$.  Let $T_{\delta'_{b+1}}$ denote tree associated with partial instance $\delta'_{b+1}$. Similarly, let $T_{\delta_{b+i}}$ be the tree associated with instance $\delta_{b+i}$. It is easy to see that $T_{\delta'_{b+1}}$ is an isomorphic sub-tree of $T_{\delta_{b+i}}$. Let $Susp(\delta'_{b+1})$ denote the set of suspended nodes in $\delta'_{b+1}$. Let $X$ denote the isomorphic nodes of $Susp(\delta'_{b+1})$ in $T_{\delta_{b+i}}$. Let $Nodes(X)$ denote the represent the set of all nodes in the sub-trees rooted for every $x \in X$. Now for each $y \in Nodes(X)$ add the rule (fact), $\{not\_modified(y)\}$ to $\Pi_{\delta_{b+i}}$.
        (Every occurence of $not \ modified(X)$ in original encoding $\Pi_{\sigma}$ can be replaced with $not\_modified(X)$ while adding another rule $\{not\_modified(X) \leftarrow not \ modified (X)\}$ without affecting correctness)
        From the rule \textit{Suspended-Unmod} $\in \Pi_{\delta'_{b+1}}$ it is clear that the nodes in $Susp(\delta'_{b+1})$ are unmodified. Similarly, for all nodes in $X$ the sub-trees rooted at nodes in $X$ are unmodified, due to the newly added rule in $\Pi_{\delta_{b+i}}$. Assume, for the sake of contradiction that $\Pi_{\delta_{b+i}}$ is unsatisfiable. Let $T'_{\delta_{b+i}}$ be the sub-tree in $\delta_{b+i}$ isomorphic to $T_{\delta'_{b+1}}$ in $\delta'_{b+1}$. Clearly, allowing modifications to just nodes in $T'_{\delta_{b+i}}$ is inadequate to satisfy $\Pi_{\delta_{b+i}}$. Then, similarly $T_{\delta'_{b+1}}$ is inadequate in $\Pi_{\delta'_{b+1}}$. But this is a contradiction. Therefore, $\Pi_{\delta_{b+i}}$ is satisfiable. The argument applies to all instances $\delta_{b+i} \in S_{\Delta}, i = 1,2,3..$. Therefore, deduction in partial instance is equivalent to deduction in complete instance. That is, Lemma 3 is true $\square$

\end{document}